\documentclass{article}

\usepackage{graphicx}
\usepackage[utf8]{inputenc}
\usepackage[T1]{fontenc}
\usepackage{cite}

\def\df{\partial}
\def\diag{\mathop{\mbox{diag}}}

\begin{document}

\title{Galactic model with a phase transition\\ from dark matter to dark energy}
\author{Igor Nikitin\\
Fraunhofer Institute for Algorithms and Scientific Computing\\
Schloss Birlinghoven, 53757 Sankt Augustin, Germany\\
\\
igor.nikitin@scai.fraunhofer.de
}
\date{}
\maketitle

\begin{abstract}
This work continues the construction of a recently proposed model of dark matter stars. In this model, dark matter quanta are sterile massless particles that are emitted from the central regions of the galaxy in the radial direction. As a result, at large distances $r$ from the center of the galaxy, the mass density of dark matter has the form $\rho \sim r^{-2}$, in contrast to the homogeneous model $\rho=Const$. In the cosmological context, the homogeneous model with massless particles corresponds to the radiation epoch of the expansion of the universe, while the proposed inhomogeneous model turns out to be equivalent to $\Lambda$CDM. In this paper, scenarios will be considered in which the radial emission of dark matter is brought into hydrostatic equilibrium with a uniform background. It is shown that solutions exist if the uniform background has an equation of state typical for dark energy. Thus, this model describes a phase transition from dark matter inside the galaxy to dark energy outside of it. The specific mechanism for such a transition could be Bose-Einstein condensation. In addition, the question of what happens if dark matter particles are not sterile, for example, are photons of the Standard Model, is considered. 
\end{abstract}


\begin{figure}
\begin{center}
\includegraphics[width=0.8\textwidth]{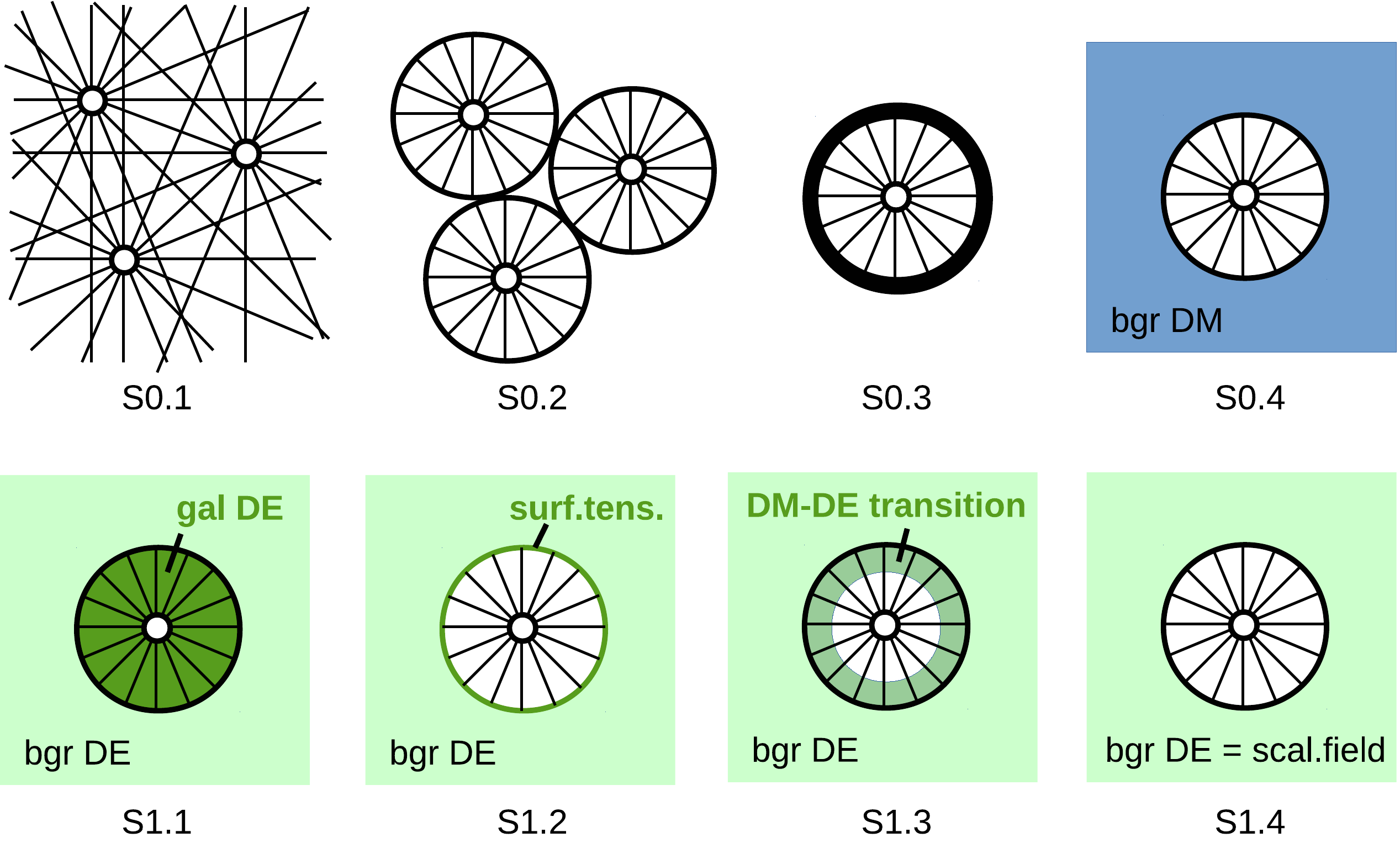}
\end{center}
\caption{The considered scenarios for a connection of galactic dark matter halo with uniform background, at the top -- rejected, at the bottom -- accepted ones.}\label{f0}
\end{figure}

\section{Introduction}
This work continues the construction of the model \cite {bled2020} presented at Bled 2020 Workshop ``What Comes Beyond the Standard Models?''. In this model, the sources of dark matter are Planck cores, Planck density objects located inside black holes. These objects are permanently emitting particles of dark matter, of originally Planck energy and Planck flux density. In this work, massless particles will be considered as quanta of dark matter, that are also sterile, which means that they do not enter into any interactions except gravitational. Radiation occurs in a T-symmetric way, into the future and into the past, so no energy is spent during the radiation, and such objects retain their mass. The radiation occurs in the radial direction; therefore, the considered flows have no transverse pressure. We denote this type of matter as null radial dark matter (NRDM).

The solution of Einstein's field equations with such matter term has a structure different from the Schwarzschild's one. This explains the designation of such compact massive objects as quasi-black holes, dark stars, Planck stars \cite {0902.0346,1612.04889,1401.6562,1409.1501}. These solutions do not have an event horizon; instead, a deep gravitational well is formed at the gravitational radius. In our model, calculations for realistic astrophysical scenarios show the redshift value $ z \sim 10 ^ {49} $, which leads to a shift of the emitted dark matter from the Planck's $ \lambda_ {in} \sim 10 ^ {- 35} $m to the ultrahigh wavelengths $ \lambda_ {out} \sim 10 ^ {14} $m, respectively, ultralow energies $ E_ {out} \sim 10 ^ {- 20} $eV. Such extreme values complicate direct detection of isolated dark matter particles. Nevertheless, the total number of emitted quanta corresponds to the initially high Planck values. The energy density and radial pressure of such radiation creates a hidden mass distribution corresponding to the observed rotation curves of galaxies. The total mass contained in such radiation, due to its extension, significantly exceeds the mass of the emitting object within its gravitational radius. The geometric mass density profile $ \rho \sim r ^ {- 2} $ typical for the radial radiation creates a linearly growing mass profile $ M \sim r $ and flat rotation curves $ v ^ 2 = GM / r = Const $. Taking into account the contributions of all black holes in the galaxy, supermassive and stellar mass ones, the distributions are modulated, and the observed nonflat rotation curves of galaxies are reproduced with good accuracy. In addition, consideration of the astrophysical scenario with the fall of an asteroid onto the Planck core leads to electromagnetic radiation with the characteristics of Fast Radio Bursts.

In this work, the main attention will be paid to the following question. If we count the massless dark matter as homogeneous (hot dark matter, HDM), then the solution of the Friedmann equations will correspond to the radiation epoch and will not coincide with the current evolution, which in the standard model corresponds to a mixture of contributions from uniform cold dark matter (CDM) and dark energy (DE). However, in the model under consideration, the distribution of matter is inhomogeneous, and, as we will see, it allows the construction of models that are in agreement with the experiment. Thus, within the framework of this model, NRDM mimics CDM at the cosmological level. The CDM macro-particles are galaxies with massive halos surrounding them. 

In more detail, we will consider several scenarios for the connection of galaxies in NRDM configuration with a uniform background. The backgrounds considered are vacuum, CDM and matter with  DE equation of state. In the first two cases, totally uniform DE contribution can be also added. The hydrostatic equilibrium of the system and the correspondence of the densities to the observed $ \Omega $-parameters will be used as selection criteria. As a result of the analysis, it turned out that of the considered scenarios, only NRDM-DE connections meet the selection criteria. Such scenarios can be interpreted as a phase transition of dark matter from the NRDM state inside galaxies to the DE state outside. The specific mechanism for such a transition can be Bose-Einstein condensation (BEC).

In addition, we will consider the question of what happens if the dark matter particles are not exactly sterile, for example, are photons of the Standard Model.

Phase transitions between dark matter and dark energy have been addressed in a number of recent works. In \cite {1907.06353}, a phase transition in a system of two scalar fields was considered, with a massive phase of dark matter condensing around galaxies, while outside one of the fields was absent, and the other turned into an exponentially rolling mode corresponding to dark energy. Conceptually, this model expands the cosmons theory \cite {hep-ph/0108266}, in which there is only one scalar field representing dark energy in the exponentially rolling mode, while its fluctuations represent dark matter. In the works \cite {1808.02472,2012.01407}, a phase transition similar to the Ising model of ferromagnetism was considered, effectively generating two cosmological constants during the evolution of the universe. In the works \cite {2008.04430,1912.02830,1911.12342,1906.01823} various scenarios of phase transitions at an early stage of the evolution of the universe, with the formation of bubbles of a new vacuum -- dark energy were considered. In these scenarios, there was a transfer and filtration of dark matter through the walls of the bubbles, which in specific calculations reproduces its present abundance. In earlier works \cite {astro-ph/0503200, gr-qc/0012094}, bubbles of a different vacuum after the phase transition were stabilized and led to the formation of massive compact objects -- dark energy stars. 

More general scenarios of the interaction of dark energy and dark matter were considered in a number of works \cite {1809.05678,1812.06854,1911.04520,1903.02370}, in the framework of the so-called $Q$-phenomenology. In this approach, the components of the dark sector are considered as two massive fluids, in which, in the absence of interaction, the energies are conserved separately. When interaction is enabled between components, energy exchange occurs, parameterized by a single scalar function $ Q $. For this function, one chooses linear dependencies of elementary densities, products of their degrees -- by analogy with the kinetics of chemical reactions, and various other model forms. The calculation results were then compared with the cosmological observables. In works \cite{2010.10823,2002.06127,1804.08558,1812.03540,1907.12551,2001.05103,1908.04281,1910.09853,2009.12620,1902.10636,1908.09843,2002.03408,1908.03324,1907.01496} the interaction of dark matter and dark energy was considered in relation to cosmological tensions. These are the discrepancies between the Hubble parameter and other cosmological properties, found in different types of observations, in particular, for the early and late stages of the evolution of the universe. The direct relation of the dark matter -- dark energy interaction models with cosmological tensions can be explained. In the absence of the interaction, the components of the dark sector evolve independently, being bound only by the common gravitational field. From here it is easy to obtain the individual dependences of the component densities on the scale factor of the universe. This makes directly observable variables (such as distribution of CMB inhomogeneities, luminosity-distance-redshift dependence, etc.) related with model parameters (such as Hubble parameter today, linear fluctuation of the matter density field, etc.). When the interaction is turned on, the components begin to pump into each other; as a result, the relationship of the model parameters with the observed variables is modified. A similar approach is used in the models of dynamical dark energy \cite {1906.09189,1701.08165,1807.03772}, where the equation of state or the density of dark energy are modified directly. The resulting changes manifest themselves as tensions between the values of the Hubble parameter, deduced from different types of measurements without model modification. Within this framework, with the right model modification, the cosmological tensions should disappear. 

The idea that dark matter and/or energy are associated with Bose-Einstein condensation, are represented by a superfluid liquid, was considered in a number of works \cite {2006.06129,1911.07371,1309.5707, hep-ph/9507385,1512.00108,1710.08910,0912.1609}. In particular, \cite {0912.1609} considered a complex scalar field with a potential equivalent to Chaplygin gas. While the specific form of the potential is not important, the presence of a minimum in it is significant. In this model, the dark energy is the state of Bose-Einstein condensate, asymptotically attained by the scalar field at this minimum. Dark matter was viewed as an excited state described by a gas of quasiparticles. In our work, a similar model will be considered, in which the outer zone of the galactic halo will also be occupied by Bose-Einstein condensate, while the distribution of dark matter in the inner part of the halo will be associated with the emission of particles from RDM stars.

First, in Section~\ref {sec2}, we will recall the structure of the RDM model, then consider a number of scenarios for its connection with a uniform background. Not all of the scenarios will be successful, but we will describe all in detail to rule out unsuccessful options. Section~\ref {sec3} considers separately the photon case. The details of the constructions are given in the Appendix. 

\section{Estimations for various scenarios}\label{sec2}

The model \cite {bled2020} considers three cases: massive, null or tachyon radial dark matter (M/N/T-RDM). The tachyon case is too exotic and will not be considered here. On the other hand, the massive case is similar to the commonly considered uniform cold dark matter (CDM). In this paper, we focus on the intermediate case, null or light-like dark matter. The quanta of such dark matter are massless sterile particles of an unspecified type.

The main formulas that determine the distribution of masses and pressures of dark matter in the model under consideration are 
\begin{eqnarray}
&\rho=p_r=\epsilon/(8\pi r^2),\ p_t=0,\ \rho_{grav}=\rho+p_r+2p_t,
\end{eqnarray}
where $ \rho $ is the mass density, $ p_r $ is the radial pressure, $ p_t $ is the transverse pressure, $ \rho_ {grav} $ is the gravitating mass density, $ r $ is the radius, $ \epsilon $ is constant scaling parameter, in the geometric system of units $ G = c = 1 $. Such dependence's are established at large distances from the center of the galaxy, when all sources of dark matter (RDM stars), distributed over the galaxy in proportion to the density of the luminous matter, can be considered as concentrated in one center. The integrated gravitating mass for such a distribution is linear in the radius: $ M_ {grav} = \epsilon r $, and the square of the orbital velocity is constant and equal to $ v ^ 2 = M_ {grav} / r = \epsilon $.

In relation to the $ \epsilon $-parameter for the Milky Way (MW) galaxy, \cite {bled2020} provides several estimates. The simplest, if one places a single RDM star in the center of the galaxy and completely neglect the contribution of the luminous matter, leads to a flat rotation curve with an orbital velocity $ v \sim200 $km/s, $ \epsilon = (v / c) ^ 2 \sim4 \cdot10 ^ {- 7} $. A more accurate estimate is obtained from the fit of the MW rotation curve, the so-called Grand Rotation Curve (GRC, \cite {VR2,0811.0859,0811.0860,1110.4431,1307.8241}). From this fit it can be seen, \cite {bled2020} Fig.2, that on an approximately flat portion of the rotation curve at the position of the Sun $ r \sim8 $kpc there is a significant contribution of luminous matter, as a result of which the contribution of dark matter to $ v ^ 2 $ is less than the trivial estimate. Further, with increasing radius, the contribution of luminous matter decreases, while the contribution of dark matter remains constant up to $ R_ {cut} \sim50 $kpc. This contribution corresponds to the galactic $ \epsilon = M_ {dm} (R_ {cut}) / R_ {cut} $, in geometric units, being averaged over the scenarios considered in \cite {bled2020}: $ M_ {dm} (R_ {cut} ) \sim2.6 \cdot10 ^ {11} M_ \odot $, $ \epsilon \sim2.5 \cdot10 ^ {- 7} $. In this work, we will carry out estimates in order of magnitude, so it is not so important which definition of the galactic $ \epsilon $ will be chosen. We prefer the latter, more precise definition and the corresponding value $ M_ {dm} (R_ {cut}) $.

When the contribution of individual black holes (identified with RDM stars) is considered, \cite {bled2020} Fig.5 gives estimates for the central supermassive and peripheral stellar black holes: $ \epsilon_ {smbh} \sim10 ^ {- 10} -10 ^ {- 7} $, $ r_ {s, smbh} \sim1.2 \cdot10 ^ {10} $m; $ \epsilon_ {sbh} \sim10 ^ {- 16} -10 ^ {- 12} $, $ r_ {s, sbh} \sim3 \cdot10 ^ 4 $m. This gives a floating estimate for the external wavelength of DM particles: $ \lambda_ {out} = r_s (8 \pi / \epsilon) ^ {1/2} $, $ \lambda_ {out, smbh} \sim10 ^ {14} -10 ^ {16} $m, $ \lambda_ {out, sbh} \sim10 ^ {11} -10 ^ {13} $m. This wavelength is highly dependent on the model used to describe the internal structure of RDM stars. In this work, for estimations, we prefer to use the value of the external wavelength $ \lambda_ {out} $ and the corresponding redshift factor $ A_ {QG} ^ {1/2} = l_P / \lambda_ {out} $ as phenomenological parameters.

The considered scenarios for the connection of galaxies in the NRDM configuration with a uniform background are schematically shown in Fig.\ref {f0}. 

\subsection{Rejected scenarios}

\paragraph {Scenario S0.1: superposition of galactic halos without cutting.} In this scenario, the dark matter halo of each galaxy extends to the radius of the visible universe $ R_ {uni} \sim14 $Gpc, dark matter from different halos does not interact and gives an additive contribution to the total mass density. If this scenario was valid, then it would be different from the radiation epoch, due to the following reasons. For the radiation epoch, the Big Bang (more precisely, the moment of recombination) is the initial flash, after which the homogeneous photon gas cools down as the universe expands. The energy of the photons changes with the scale factor as $ a ^ {- 1} $, and their numerical density as $ a ^ {- 3} $, which gives the dependence $ a ^ {- 4} $ for the mass density. For the RDM model, despite the expansion of the universe and the separation of RDM stars from each other, the energy of DM particles near RDM stars is fixed, related to the above-mentioned parameter $ \lambda_ {out} $. It is important that this energy does not fall over time. The number density of RDM stars falls as $ a ^ {- 3} $, and as a result the average mass density also falls as $ a ^ {- 3} $, just like for CDM.

The scenario is prohibited due to the following evaluation. According to calculations for the Milky Way galaxy \cite {bled2020}, the cutoff radius and halo mass are $ R_ {cut} \sim50 $kpc, $ M_ {dm} (R_ {cut}) \sim2.6 \cdot10 ^ {11} M_ \odot $, and the mass of the disk and other emitting structures can be neglected in the order estimate. If one does not use the cutoff and continues the halo to the border of the universe, $ M_ {dm} (R_ {uni}) = M_ {dm} (R_ {cut}) R_ {uni} / R_ {cut} \sim7.3 \cdot10 ^ { 16} M_ \odot $. If the result is multiplied by the estimated number of galaxies in the universe $ N_ {gal} \sim2 \cdot10 ^ {12} $, we get $ M_ {dm} \sim1.5 \cdot10 ^ {29} M_ \odot $. Compared to the estimated mass of dark matter in the universe $ M_ {dm, uni} \sim4.5 \cdot10 ^ {23} M_ \odot $, this value is overestimated by the factor $ \sim3.2 \cdot10 ^ 5 $. The mass density averaged over the volume of the universe for the obtained value of $ M_ {dm} $ will be $ \rho_ {dm} \sim8.6 \cdot10 ^ {- 22} kg / m ^ 3 $, which in $ \sim3.2 \cdot10 ^ 5 $ more than the estimate from the critical density $ \Omega_ {dm} \rho_ {crit} = 2.7 \cdot10 ^ {- 27} kg / m ^ 3 $.

The following corrections can be made to this calculation. The geometric cutoff by $ R_ {uni} $ for the galaxies located far from the center (which we locate in the MW) can reduce the halo mass, but the factor is small, at most $ 2 $. The mass density is everywhere understood as the gravitating mass density, which also includes the pressure $ \rho_ {grav} = \rho + p_r $. At cosmological distances, the energy of DM particles is redshifted, but considering distances up to $ 0.03R_ {uni} \sim420 $Mpc, the mass density decrease factor will not exceed 20\% (for the Hubble parameter $ H_0 = 72 $km/s/Mpc such distances correspond to $ z \sim0.1 $, 10\% decrease in energy and 10\% slowdown in time, affecting 20\% decrease in flux density). This can be used to estimate the mass from below, as a result the $> 7.6 \cdot10 ^ 3 $ discrepancy will remain unexplained.

Further, the estimate is based on the assumption that all galaxies have masses of the order of the MW. This, of course, is not the case, there is a distribution of galaxies by masses. An accurate account of the distribution of galaxies will be given in the Appendix, and a similar result will be obtained, within the accuracy of modeling the distribution of galaxies.

Now we assume that all galaxies are copies of the MW, the halo of each galaxy is extended to the radius $ R_ {gal} $ and the relation $ N_ {gal} M_ {dm} (R_ {cut}) / R_ {cut} \cdot R_ {gal} = M_ {dm, uni} $ holds, at which all the necessary mass relations are joined. Substituting the known data, we get $ R_ {gal} = 44 $kpc, which almost coincides with the value of $ R_ {cut, MW} $. An exact match of $ R_ {gal} = R_ {cut, MW} $ can be achieved by slightly adjusting the estimated number of galaxies to $ N '_ {gal} = 1.7 \cdot10 ^ {12} $. Thus, according to this estimate, if we imagine that the universe consists of copies of the MW galaxy, the halo of which is cut off by $ R_ {cut} \sim50 $kpc, then the total mass of dark matter in the universe will coincide in order of magnitude with its cosmological estimate. With this configuration, dark matter is entirely concentrated in galaxies and is absent in the intergalactic environment.

Let's find out what happens if the $ R_ {cut} $ parameter is increased to 1Mpc. This is possible when the estimated number of galaxies is reduced to $ N '_ {gal} \sim8.7 \cdot10 ^ {10} $. It is known that the mass-to-light ratio of galaxies stops changing at distances of this order of magnitude, see \cite {Dodelson} Fig.2.5. Moreover, this value is of the order of intergalactic distances. Thus, a scenario is theoretically possible in which galaxies touch each other in the outer zones of their halos, although it may require tensions of the $ N_ {gal} $ parameter.

Finally, for the original scenario in which the halos overlap and reach the size of the universe, it would be $ N '_ {gal} \sim6.2 \cdot10 ^ {6} $, a too strong deviation from the observed value. Therefore, this scenario can be considered as excluded. 

\paragraph {Scenario S0.2: touching halos in dynamic equilibrium.} The above variant with halos touching each other in the outer region, with the refinement that galaxies can exchange dark matter. Null dark matter leaking from one galaxy is absorbed by neighboring galaxies, and vice versa. In fact, the world lines of dark matter form a network connecting the galaxies, and the concept of a spherical halo is only an approximation.

The problem here is as follows. As a result of the expansion of galaxies, DM particles coming from neighboring galaxies are subject to a small redshift $ z $ and decrease their energy and flux density by the corresponding factor. We consider RDM stars in a stationary T-symmetric scenario. The parameters of dark matter, in particular, its energy and flux density, coincide for the incoming and outgoing flows. Therefore, the exiting particles also have a reduced energy and flux density. With multiple reflections between galaxies, the redshift of DM particles accumulates, exactly as it would in a homogeneous environment. RDM stars act as spherical mirrors, changing the direction of the DM particles, but not their energy characteristics. Such an environment turns out to be equivalent to HDM, its evolution coincides with the era of radiation, which is different from the observed evolution of the universe today.

In fact, the stationary state of RDM stars requires T-symmetry only for the energy flux density $ \epsilon $, the individual energies of incoming and outgoing DM particles can be different, compensated by different number flux densities. This will not help, since it is energy density that governs cosmological evolution. Note also that in the equation $ \rho_P = \epsilon / (8 \pi r_s ^ 2 A_ {QG}) $, which determines conditions on the surface of the Planck core, in the considered scenario the factors $ \epsilon $ and $ A_ {QG} $ are scaled equally so that the gravitational radius $ r_s $ can remain unchanged. Strictly speaking, changing $ \epsilon $ and $ A_ {QG} $ in stationary scenarios is also unacceptable, but we consider this change as performed rather slowly, quasi-statically. What happens in fast scenarios, as well as with T-asymmetric $ \epsilon $ and variable $ r_s $, can be found out only after solving the dynamic RDM problem, which goes beyond the scope of this work.

\paragraph {Scenario S0.3: a halo surrounded by a massive thin shell.} We now look at a few scenarios from the {\it termination shock} type. This phenomenon occurs at the edge of the solar system when the radially directed solar wind meets the isotropic interstellar medium. Similar phenomena can occur with dark matter at the edge of the galaxy when the radial flow of dark matter meets the intergalactic environment. First, we will consider a scenario in which an NRDM galaxy at radius $ R_ {cut} $ is surrounded by a thin CDM layer, and there is a vacuum outside. The CDM layer is held in equilibrium by the NRDM pressure force and the force of gravity. If such a scenario was possible, the galaxies would be isolated from each other and would be massive balls floating in a vacuum. On a cosmological level, such matter is equivalent to CDM.

The equilibrium condition of forces can be written as $ \epsilon / (8 \pi r ^ 2) \cdot 4 \pi r ^ 2 = \epsilon / r \cdot m $ for $ r = R_ {cut} $, whence the mass of the CDM layer $ m = R_ {cut} / 2 $, in geometric units. This is a huge mass, exceeding the mass of the galaxy $ M_ {dm} (R_ {cut}) = \epsilon R_ {cut} $, where $ \epsilon \ll1 $, for MW $ \epsilon = 2.5 \cdot10 ^ {- 7} $. Formally, with such a mass, the galaxy is covered by its event horizon, becoming a black hole. More precisely, the calculation uses Newtonian equations and only shows that there is no solution in weak fields. The interpretation of this result is that the relativistic pressure at the boundary of the NRDM galaxy can be compensated only by relativistic gravitational forces.

The distribution of matter in the CDM layer obeys the Tolman-Oppenheimer-Volkoff (TOV) equation, the solution of which in weak fields and for a thin layer is described by the one-dimensional hydrostatic equation $ \rho = \rho_0 \exp (-gh / w) $, where $ w $ - parameter of the equation of state (EOS) $ p = w \rho $, for CDM $ w = kT / m \ll1 $, all equations are written in geometric units. The pressure equilibrium at the boundary of the layer leads to $ \epsilon / (8 \pi R_ {cut} ^ 2) = w \rho_0 $, also $ g = \epsilon / R_ {cut} $, whence $ \rho = \epsilon / ( 8 \pi R_ {cut} ^ 2 w) \cdot \exp (- \epsilon h / (wR_ {cut})) $. Integrating this value, we get $ m = 4 \pi R_ {cut} ^ 2 \int \rho dh = R_ {cut} / 2 $. The result is independent of $ w $ and coincides with the estimate above. 

\paragraph {Scenario S0.4: halo surrounded by homogeneous dark matter.} A variation of the previous scenario, where instead of vacuum there is a homogeneous dark matter with isotropic EOS outside: $ p_ {bgr} = w \rho_ {bgr} $. Here we will consider two options: CDM $ w \ll1 $, HDM $ w = 1/3 $. Pressure equilibrium at the halo boundary: $ \epsilon / (8 \pi R_ {cut} ^ 2) = w \rho_ {bgr} $, gravitating masses: $ M_ {dm, gal} = N_ {gal} \epsilon R_ {cut } $, $ M_ {dm, bgr} = (1 + 3w) \rho_ {bgr} \cdot (4 \pi / 3) (R_ {uni} ^ 3-N_ {gal} R_ {cut} ^ 3) $, an estimate of the total mass of dark matter in the universe: $ M_ {dm, uni} = M_ {dm, gal} + M_ {dm, bgr} = N_ {gal} \epsilon R_ {cut} + \epsilon (1 + 3w) / (6w R_ {cut} ^ 2) (R_ {uni} ^ 3-N_ {gal} R_ {cut} ^ 3) $. Note that, according to earlier calculations, the first term already corresponds in order to the cosmological estimate for the mass of dark matter. Only if the second term is small, this correspondence could be preserved. However, if we assume that the intergalactic distances significantly exceed the size of the halo, and the estimate $ R_ {uni} ^ 3 \gg N_ {gal} R_ {cut} ^ 3 $ holds, then the second term $ M_ {dm, uni} \sim \epsilon R_ {uni} ^ 3 / R_ {cut} ^ 2 \cdot (1 + 3w) / (6w) $, which for $ \epsilon = 2.5 \cdot10 ^ {- 7} $, $ R_ {uni} \sim14 $Gpc and $ R_ {cut} \sim50 $kpc matches $ M_ {dm, uni} /M_\odot\sim5.7\cdot10^ {27} (1 + 3w) / (6w) $. It can be seen that already for $ w = 1/3 $ and even more so for $ w \ll1 $ the result significantly exceeds the value $ M_ {dm, uni} /M_\odot\sim4.5\cdot10^ {23} $, obtained from cosmological estimates. Also for the above option with tension, $ R_ {cut} = 1 $Mpc, $ N '_ {gal} = 8.7 \cdot10 ^ {10} $ the resulting formula is $ M_ {dm, uni} /M_\odot=4.5 \cdot10 ^ {23} +1.4 \cdot10 ^ {25} (1 + 3w) / (6w) $ does not allow CDM/HDM as background matter, for continuous matching with NRDM pressure at halo boundaries. 

\subsection{Accepted scenarios}

Next, we'll look at scenarios involving dark energy. We will represent dark energy as a kind of matter, perhaps a kind of dark matter or its other phase state, which has an isotropic EOS $ p_ {de} = - \rho_ {de} $, that is, $ w = -1 $, with positive $ \rho_ {de} $, constant within each phase. The density of the gravitating mass for such matter is negative and is equal to $ \rho_ {de, grav} = \rho_ {de} + 3p_ {de} = - 2 \rho_ {de} $. The negativity of this density, provided that it prevails over other components, is the driving mechanism for the accelerated expansion of the universe.

\paragraph {Scenario S1.1: jump in the dark energy density at the halo boundary.} Let there be two different dark energy densities, outside the halo $ \rho_ {de, bgr} $, inside the halo $ \rho_ {de, gal} $, with a jump at $ R_ {cut} $. Equilibrium pressure condition $ \epsilon / (8 \pi R_ {cut} ^ 2) = p_ {de, bgr} -p_ {de, gal} = \rho_ {de, gal} - \rho_ {de, bgr} $, gravitating masses: $ M_ {dm, gal} = N_ {gal} \epsilon R_ {cut} $, $ M_ {de, gal} = - 2 \rho_ {de, gal} N_ {gal} \cdot (4 \pi / 3) R_ {cut} ^ 3 $, $ M_ {de, bgr} = - 2 \rho_ {de, bgr} \cdot (4 \pi / 3) (R_ {uni} ^ 3-N_ {gal} R_ {cut} ^ 3) $, an estimate of the total mass of dark matter and dark energy in the universe: $ M_ {dm + de, uni} = M_ {dm, gal} + M_ {de, gal} + M_ {de, bgr} = (2/3) \epsilon N_ {gal} R_ {cut} - (8 \pi / 3) \rho_ {de, bgr} R_ {uni} ^ 3 $. The second term here describes the total gravitating mass of dark energy, as if it uniformly filled the entire universe, including galactic halos. The first term is the gravitating mass of the galactic halo reduced by the factor $ (2/3) $. In general, the model behaves like a mixture of uniform cold dark matter and uniform dark energy, like $ \Lambda $CDM. In order of magnitude, for $ R_ {cut} = 50 $kpc, the CDM mass corresponds to cosmological estimates. Fine tuning is also possible similar to scenario S0.1, the factor $ (2/3) $ can be compensated for by a small increase in the estimated number of galaxies $ N '_ {gal} = 2.6 \cdot10 ^ {12} $.

Let us also analyze the expression for the gravitating mass of one galaxy: $ M (r) = \epsilon r -2 \rho_ {de, gal} (4 \pi / 3) r ^ 3 $. In the expression for the internal dark energy density $ \rho_ {de, gal} = \epsilon / (8 \pi R_ {cut} ^ 2) + \rho_ {de, bgr} $ for the selected value $ R_ {cut} = 50 $kpc, after conversion to natural units, the first term is $ 5.6 \cdot10 ^ {- 24} kg / m ^ 3 $, the second $ \rho_ {de, bgr} = \Omega_ {de} \rho_ {crit} = 6.8 \cdot10 ^ {- 27} kg / m ^ 3 $, the first term dominates. Thus, continuous matching of pressures at the galactic boundary requires a jump in the dark energy density by a factor of $ \sim10 ^ 3 $. Note that this jump can be reduced by adjusting the $ R_ {cut} $ parameter.

Further, the expression for the mass function at the selected parameters becomes $ M (r) /M_\odot=2.6\cdot10^ {11} (r / R_ {cut}) - 8.7 \cdot10 ^ {10} (r / R_ {cut} ) ^ 3 $. In the inner part of the rotation curve, for example, up to the position of the Sun $ r \sim8 $kpc, the first term dominates. Thus, the interior of the rotation curve is unaffected by the dark energy introduced into the model. In the outer part of the curve, the contribution of the enhanced internal dark energy density becomes noticeable, finally, it is this contribution that leads to the factor $ (2/3) $ in the mass formulas. The term proportional to the external dark energy density for the chosen parameters of the model makes a negligible contribution within the galaxy. It begins to dominate in the formula $ M (r> R_ {cut}) = (2/3) \epsilon R_ {cut} - (8 \pi / 3) \rho_ {de, bgr} r ^ 3 $ at distances $ r > 0.6 $Mpc, at which the effects of cosmological expansion become noticeable.

It becomes clear that in the considered scenario the rotation curve undergoes a change only in its outer part, where it decreases by $ \sqrt {2/3} $ factor, about $ 18 \% $. As we will see later, the measurement errors in this range significantly exceed this variation, which makes it impossible to distinguish this solution from the reference profile.

Thus, we have obtained the first scenario, which connects null matter in galactic halos with a cosmological background of dark energy and turns out to be equivalent to the uniform $ \Lambda $CDM model. A calculation based on a simple equilibrium of pressures does not provide any indication for the possible nature of the increased density of dark energy within the galaxy. Phenomenologically, dark energy can be described as a medium in which its constituent particles experience mutual attraction. This attraction corresponds to negative pressure, while the work of external forces $ -pdV $ is used to increase the internal energy $ \rho dV $, in accordance with EOS $ -p = \rho $. The presence of two phases with different pressures suggests two varieties for such media. An analogy can be drawn here with the string model. The energy of a string is proportional to its length, just like the total mass for dark energy is proportional to its volume. The strings have a fixed tension, which is a constant in the model. One can consider strings with different tensions as separate varieties of the same model. The considered scenario demonstrates a fundamental possibility; further possible alternatives will be considered.

\paragraph {Scenario S1.2: surface tension at the boundary between the halo and the background from dark energy.} Let inside $ R_ {cut} $ be NRDM, outside -- dark energy with density $ \rho_ {de, bgr} $, and surface tension with coefficient $ \sigma $ acts on the boundary. Equilibrium pressure condition $ \epsilon / (8 \pi R_ {cut} ^ 2) = 2 \sigma / R_ {cut} + p_ {de, bgr} = 2 \sigma / R_ {cut} - \rho_ {de, bgr } $, gravitating masses: $ M_ {dm, gal} = N_ {gal} \epsilon R_ {cut} $, $ M_ {de, surf} = - N_ {gal} \sigma \cdot 4 \pi R_ {cut} ^ 2 $, $ M_ {de, bgr} = - 2 \rho_ {de, bgr} \cdot (4 \pi / 3) (R_ {uni} ^ 3-N_ {gal} R_ {cut} ^ 3) $, an estimate of the total mass of dark matter and dark energy in the universe: $ M_ {dm + de, uni} = M_ {dm, gal} + M_ {de, surf} + M_ {de, bgr} = (3/4) \epsilon N_ {gal} R_ {cut} + (2 \pi / 3) \cdot N_ {gal} R_ {cut} ^ 3 \rho_ {de, bgr} - (8 \pi / 3) \rho_ {de, bgr} R_ {uni} ^ 3 $. Here the third term corresponds to the cosmological contribution of dark energy, it grows in the negative direction in proportion to the volume of the expanding universe. The first and second terms are preserved in the expansion and represent CDM. At $ R_ {cut} = 50 $kpc, the first term significantly exceeds the second, and, as in the previous scenario, allows fine tuning of the parameters to the cosmological value of CDM density.

In the above formulas, the gravitating mass corresponding to the boundary layer is calculated as follows. Surface tension is related to negative transverse pressure and positive energy density as $ -p_t = \rho = \sigma / dr $, where $ dr $ is the layer thickness. The gravitating mass of the spherical layer is $ M = (\rho + 2p_t) Sdr = - \sigma \cdot 4 \pi R_ {cut} ^ 2 $. There is also a radial pressure $ p_r $ inside the layer, which continuously interpolates the boundary values, remains bounded, and makes a vanishing contribution at $ dr \to0 $.

When choosing $ R_ {cut} = 50 $kpc, the density jump between external dark matter and NRDM is still $ \sim10 ^ 3 $ times, but here it is compensated by surface tension. As in the previous scenario, the pressure jump can be reduced by adjusting the $ R_ {cut} $ parameter. The mass function for $ r <R_ {cut} $ coincides with the NRDM dependence $ M (r <R_ {cut}) = \epsilon r $, thus the inner rotation curve does not change. When passing through $ R_ {cut} $, the mass function undergoes a jump $ M (R_ {cut} +0) = (3/4) \epsilon R_ {cut} - 2 \pi R_ {cut} ^ 3 \rho_ {de, bgr} $, the coefficient $ (3/4) $ appears in the first term, and the second term also appears. With the chosen parameters, the first term is $ 1.9 \cdot10 ^ {11} M _ {\odot} $, the second $ -7.9 \cdot10 ^ 7M _ {\odot} $, the first term dominates. Further, the mass function includes the cosmological term $ M (r> R_ {cut}) = (3/4) \epsilon R_ {cut} + (2 \pi / 3) R_ {cut} ^ 3 \rho_ {de, bgr} - (8 \pi / 3) \rho_ {de, bgr} r ^ 3 $, which dominates for $ r> 0.6 $Mpc.

The resulting scenario is very close to the previous one, only a different mechanism is used to compensate for the pressure jump at the edge of the galaxy. Phenomenologically, if we consider dark energy as a medium consisting of interacting particles, the presence of a boundary can lead to the appearance of a surface term in the equations, as for classical media. The jump in the mass function that appears in this scenario corresponds to a jump in the rotation curve by the factor $ \sqrt {3/4} $, about 13\%. This jump also occurs in the outer region, where the scatter of experimental data is large, so that it can be unnoticed. Also, this jump can be an idealization of a more complex scenario in which the transition layer has a finite thickness. The possibility of a gradual change of EOS will be explored in the following scenario. 

\paragraph {Scenario S1.3: phase transition of dark matter to dark energy.} In this scenario, we assume that dark energy is a form of dark matter, and with the increasing radius, there is a continuous transition between the corresponding EOS: $ p_r = w_r \rho $, $ p_t = w_t \rho $, $ (w_r, w_t) $ change from $ (1,0) $ for $ r = R_ {cut1} $ to $ (- 1, -1) $ for $ r = R_ {cut2 }> R_ {cut1} $. The result depends on the transition path, which we fix from physical considerations as follows. Initially, from $ r = R_ {cut1} $ to the intermediate point $ r = R_ {cut1b} $, only $ w_t $ changes, from $ 0 $ to $ -1 $. The inclusion of transverse attraction between flows of dark matter leads to the Joule-Thomson effect known in gas dynamics, cooling of flows, which in our case manifests itself in a rapid decrease in the mass density $ \rho $. Further, from $ r = R_ {cut1b} $ to $ r = R_ {cut2} $ only $ w_r $ changes, from $ 1 $ to $ -1 $. In this range, the contributions of dark matter from different sources are mixed, the matter becomes isotropic. Further, the matter obeys isotropic EOS for dark energy, and its density and pressure become constant.

It is convenient to solve the problem in logarithmic variables $ x = \log r $, $ \xi = \log \rho $, with the restriction $ \rho> 0 $. To interpolate $ w_ {t, r} $ in the corresponding intervals, we choose functions linear in $ x $, and the positions of the endpoints $ \{x_1, x_ {1b}, x_2 \} = \log \{R_ {cut1}, R_ {cut1b}, R_ {cut2} \} $ will be chosen from the correspondence of the model to the cosmological parameters.

The stationary spherically symmetric solutions considered here satisfy the hydrostatic equation for anisotropic medium, see Appendix for details: $ r (p_r + \rho) A'_r + 2A (r (p_r) '_ r + 2p_r-2p_t) = 0 $. The first term describes the gravitational interaction, which in our problems can be neglected. The reason for this is that in the weak field limit $ A \sim1 + 2 \phi $, $ A'_r \sim2g $, where $ \phi $ is the gravitational potential, $ g = \phi'_r = M_ {grav } (r) / r ^ 2 $ is gravitational field in the used system of units, $ | \phi | \ll1 $, $ rg \ll1 $. In our models, the density and pressure are controlled by a small common factor $ \epsilon $, and the first term turns out to be of the next smallness order compared to the second one. This property of the weak-field regime can also be verified on the exact solutions of the hydrostatic equation, given in Appendix.

Thus, we can concentrate on the second term: $ r (p_r) '_ r + 2p_r-2p_t = 0 $. Let's go to logarithmic variables and substitute EOS: $ w_r \xi'_x + (w_r) '_ x + 2 (w_r-w_t) = 0 $. The solution is: $ \xi = - \int dx ((w_r) '_ x + 2 (w_r-w_t)) / w_r $. In the following, we will consider regular solutions in which the denominator and the numerator in the integrand vanish simultaneously: $ w_r = 0 $, $ (w_r) '_ x = 2w_t $. Note that, with our choice of the interpolation order, the condition $ w_r = 0 $ can be satisfied only at the second stage, in the interval $ [x_ {1b}, x_2] $, while, due to the linearity of the interpolation, the condition $ (w_r) '_ x = 2w_t = -2 $ holds on this entire interval.

At the first stage $ [x_1, x_ {1b}] $, $ (w_r, w_t) = (1, -q) $, $ q = (x-x_1) / (x_ {1b} -x_1) $, calculating the integral, we get $ \xi_ {1b} - \xi_1 = -3 (x_ {1b} -x_1) $. At the second stage $ [x_ {1b}, x_ {2}] $, $ (w_r, w_t) = (1-2q, -1) $, $ q = (x-x_ {1b}) / (x_2-x_ {1b}) $, from the condition $ (w_r) '_ x = -2 $ we get $ q'_x = 1 $, that is, $ x_2 = x_ {1b} + 1 $. Calculating the integral, we get $ \xi_2- \xi_ {1b} = - 2 $. Hence $ \log (\rho_1 / \rho_2) = \xi_1- \xi_2 = 3 (x_ {1b} -x_1) + 2 $. Choosing $ \rho_1 = \epsilon / (8 \pi R_ {cut1} ^ 2) $, $ R_ {cut1} = R_ {cut} = 50 $kpc, $ \epsilon = 2.5 \cdot10 ^ {- 7} $, $ \rho_2 = \rho_ {de, bgr} = \Omega_ {de} \rho_ {crit} = 6.8 \cdot10 ^ {- 27} kg / m ^ 3 $, and also converting all values into the natural system of units, we get: $ \rho_1 / \rho_2 = 824 $, $ R_ {cut1b} = 0.24 $Mpc, $ R_ {cut2} = 0.65 $Mpc. Thus, the required density variation from NRDM to the background dark energy in the considered scenario fixes the halo cutoff parameters to reasonable values.

Next, consider the contribution of the galaxy to the cosmological mass density. The gravitating mass density is $ \rho_ {grav} = (1 + w_r + 2w_t) \rho $, and the gravitating mass of the spherical layer is $ \Delta M_ {grav} = 4 \pi \int \rho_ {grav} r ^ 2dr $. After the transition to logarithmic variables, the integrals over two interpolation intervals can be taken analytically. Omitting cumbersome expressions, we will immediately give the numerical answer $ \{M_1, \Delta M_1, \Delta M_2, M_ {vac} \} = \{2.60, 2.67, -2.60, 2.35 \} \cdot10 ^ {11} M_ \odot $. Here $ M_1 = \epsilon R_ {cut} $ is the mass of the NRDM halo, $ \Delta M_ {1,2} $ are the masses of the spherical layers for two interpolation intervals, $ M_ {vac} = (8 \pi / 3) \rho_ {de, bgr} R_ {cut2} ^ 3 $ is the compensation mass of the vacuole arising from the rearrangement of the terms $ M_ {dm + de, uni} = N_ {gal} M_ {dm + de, gal} - (8 \pi / 3) \rho_ {de, bgr} (R_ {uni} ^ 3-N_ {gal} R_ {cut2} ^ 3) = N_ {gal} (M_ {dm + de, gal} + M_ {vac}) - (8 \pi / 3) \rho_ {de, bgr} R_ {uni} ^ 3 $. The $ M_ {vac} $ term should be taken into account in cosmological calculations, when reducing to the parameters of a homogeneous medium, while when calculating the rotation curves only the actually present masses should be taken, and $ M_ {vac} $ should be omitted. Interestingly, there is an identity $ M_1 + \Delta M_2 = 0 $, which holds exactly, at the analytical level, but is probably a coincidence due to a special choice of interpolating functions. Also of interest is the approximate equality of all mass contributions in their absolute value. The cosmological mass per galaxy is the sum of all these contributions and is equal to $ M_ {dm + de, gal} + M_ {vac} = 5 \cdot10 ^ {11} M_ \odot $. This gives a coincidence with the cosmological CDM mass $ M_ {dm, uni} = 4.5 \cdot10 ^ {23} M_ \odot $ in order of magnitude, for exact coincidence the estimated number of galaxies should be reduced to $ N '_ {gal} = 9 \cdot10 ^ {11} $, 2.2 times less than the nominal value. One can also adjust the $ \epsilon $ parameter, but since our estimates of the halo cutoff parameters were tied to MW values, these estimates must be repeated when $ \epsilon $ changes.

The constructed scenario, obviously, contains wide arbitrariness in the choice of interpolating functions and is rather a proof of the existence of a solution satisfying cosmological estimates. This existence in itself is non-trivial. Recall that in standard cosmology, null, hot dark matter leads to a different rate of cosmological expansion today and is forbidden. The possibility of joining hot dark matter with dark energy within the galactic halo, at a cosmological level equivalent to $ \Lambda $CDM, is the main result of this work. The specific way of joining may be different, in the Appendix we will discuss the possibility of narrowing this arbitrariness.

For now, note that the interpolation order selected in the model is significant. The reverse order when $ (w_r, w_t) $ changes from $ (1,0) $ to $ (- 1,0) $ for $ r \in [R_ {cut1}, R_ {cut1b}] $ leads to the condition $ (w_r) '_ x = 2w_t = 0 $, not feasible for linear functions. If we interpolate both terms at the same time, $ (w_r, w_t) = (1-2q, -q) $, $ q = (x-x_1) / (x_ {2b} -x_1) $, from the conditions $ w_r = 0 $ , $ (w_r) '_ x = 2w_t $ we get $ q = 1/2 $, $ q'_x = 1/2 $, that is, $ x_ {2b} = x_1 + 2 $. Moreover, $ \xi_ {2b} = \xi_1-2 $, which for $ R_ {cut1} = 50 $kpc gives $ R_ {cut2b} = e ^ 2R_ {cut1} = 0.37 $Mpc, $ \rho_1 / \rho_ {2b} = e ^ 2 \sim7.4 $, far from the experimental value of $ \rho_1 / \rho_2 \sim824 $. The physical rationale with the initial cooling of dark matter due to the Joule-Thomson effect and the subsequent transition to the isotropic phase for the cooled gas was important for obtaining the strong density drop observed in real galaxies.

\begin{figure}
\begin{center}
\includegraphics[width=\textwidth]{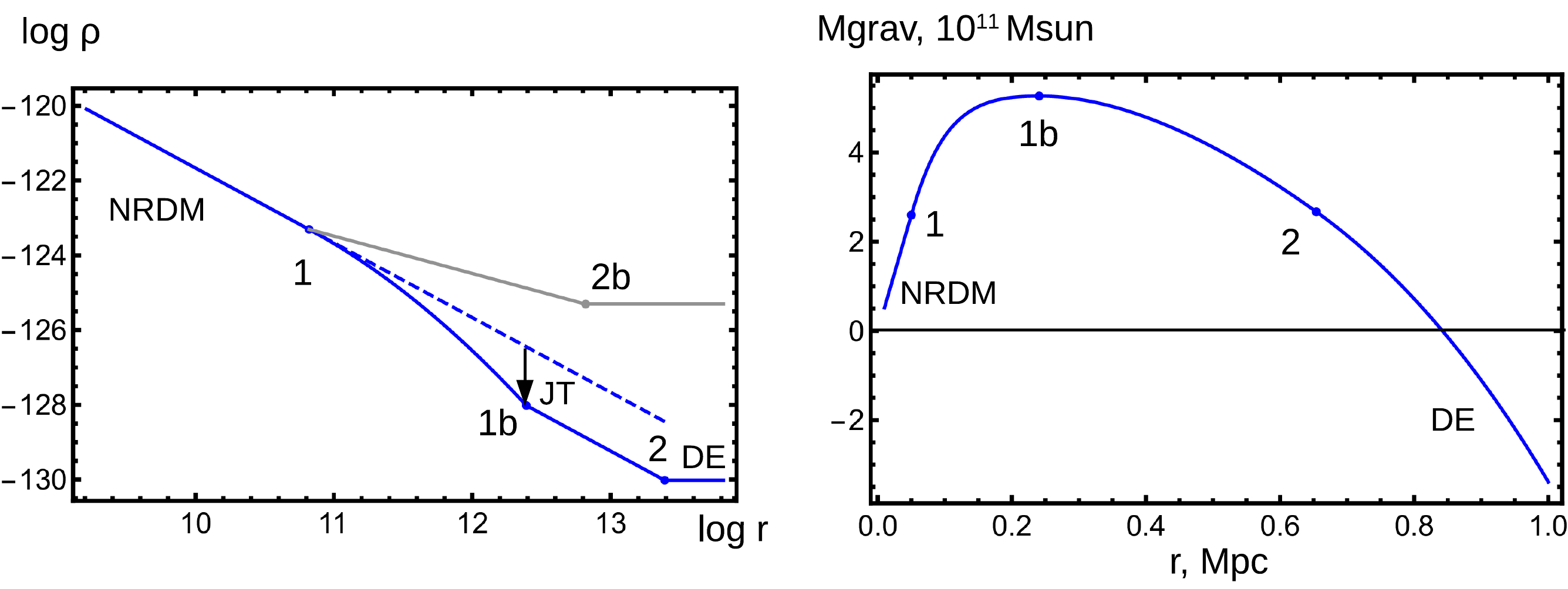}
\end{center}
\caption{Physical profiles for scenario S1.3.}\label{f1}
\end{figure}

Here are some graphs showing the behavior of the main physical profiles in the considered scenario. Fig.\ref {f1} left shows the dependence of $ \xi = \log \rho $ on $ x = \log r $. Initially, the graph contains an NRDM line with a slope of $ -2 $, which corresponds to the $ \rho \sim r ^ {- 2} $ dependence. Further, at point 1, the transverse interaction between the flows turns on, and the Joule-Thomson effect is superimposed on the continuing radial drop in density. Here, the slope of the graph $ d \xi / dx $ is continuously changing from $ -2 $ to $ -4 $. Further, in the interval from 1b to 2, a transition to the isotropic phase follows, the slope in this case being equal to $ -2 $. After point 2, there is isotropic dark energy with constant density, slope $ 0 $. The resulting density variation between points 1 and 2 corresponds to the experimentally observed factor of $ \rho_1 / \rho_2 \sim824 $.

For comparison, the option shown in gray when $ (w_r, w_t) $ are linearly interpolated at the same time. The slope between points 1 and 2b is $ -1 $. After 2b, there is an isotropic phase with a slope of $ 0 $. Due to these changes, the graph goes much higher than the previous one, the density variation does not correspond to the observed value.

Fig.\ref {f1} right shows the dependence of $ M_ {grav} (r) $. Initially, there is an NRDM part with a characteristic linear dependence, then at point 1b, the dependence passes through a maximum and, after point 2, is described by a negative cubic term corresponding to the contribution of dark energy. 

\begin{figure}
\begin{center}
\includegraphics[width=\textwidth]{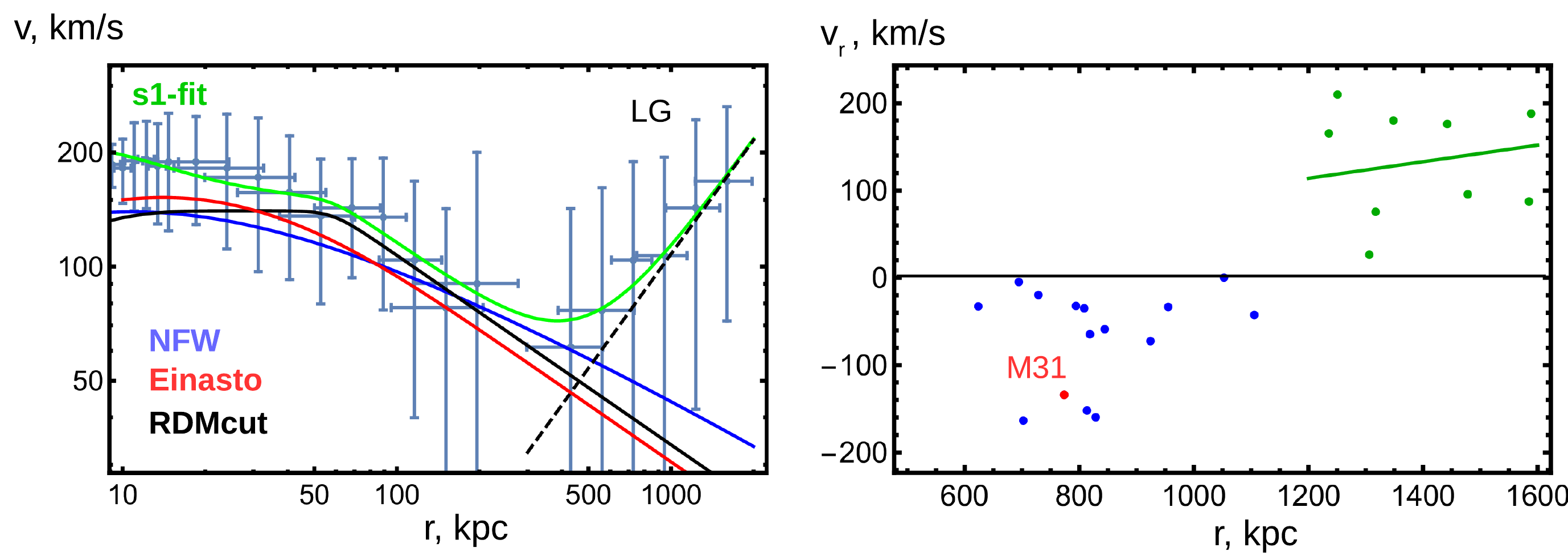}
\end{center}
\caption{Left: an external part of the Milky Way rotation curve, according to \cite {1307.8241}. A variety of profiles are shown, including the RDMcut scenario from \cite {bled2020}. Right: outer part of the dependence of radial velocity on the distance, according to \cite {0811.0860}. The position of the galaxy M31 is marked, the outer part of the graph is fitted with a Hubble-alike dependence.}\label{f2}
\end{figure}

\begin{figure}
\begin{center}
\includegraphics[width=\textwidth]{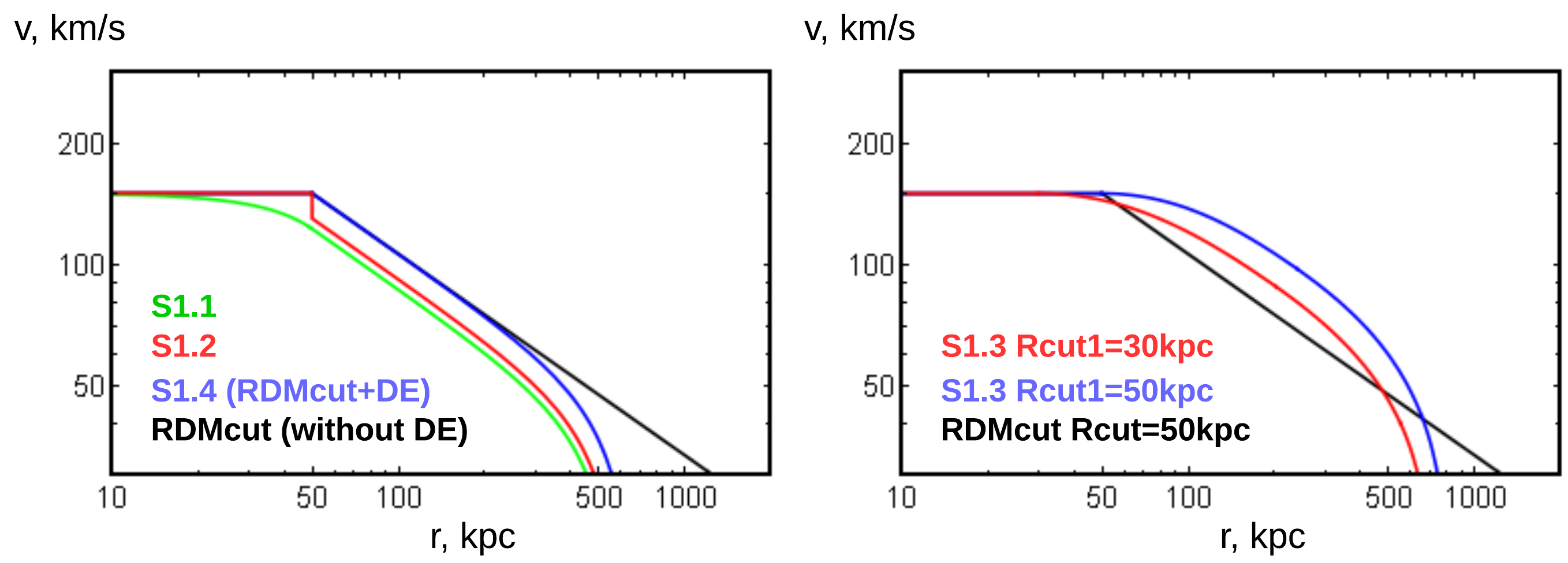}
\end{center}
\caption{Profiles built in the scenarios of this work, compared with the RDMcut profile.}\label{f3}
\end{figure}

\paragraph*{Scenario S1.4: Bose-Einstein condensation.} In this scenario, two phases are also considered: the internal NRDM phase, described by the classical particle model, and the external phase, described by a complex scalar field. This field theory is used in phenomenological models of Bose-Einstein condensation, as well as in cosmological models of quintessence and its variants (k-essence, quartessence, Chaplygin gas), see \cite {0912.1609} and references therein. Therefore, this scenario assumes that dark matter particles are emitted by RDM stars in the galaxy and undergo Bose-Einstein condensation at large distances. Alternatively, these can be particles of different types that are in contact equilibrium at the edge of the galactic halo.

In the field theory under consideration, the Lagrangian, the energy-momentum tensor, and the equations of motion have the form \cite {Blau} Chap.6.3,7.5: 
\begin{eqnarray}
&&L=-(\df_\mu \phi^* \df^\mu \phi)/2 - V(|\phi|^2),\\
&&T_{\mu\nu}=(\df_\mu \phi^* \df_\nu \phi + \df_\nu \phi^* \df_\mu \phi)/2+g_{\mu\nu}L,\\
&&(-\df^2/\df t^2+\Delta)\phi=2V'(|\phi|^2)\phi.
\end{eqnarray}
Here the equations of motion are written in a flat background, and the rest of the expressions are valid for an arbitrary metric. We also remind that for a scalar field the covariant and coordinate derivatives are equal: $ \nabla_ \mu \phi = \df_ \mu \phi $. The field equations belong to the well-known nonlinear Klein-Gordon type with the potential. For $ V (| \phi | ^ 2) = Const + m ^ 2 | \phi | ^ 2/2 $ the equations become linear and describe the behavior of a free massive scalar field. We neglect the influence of gravity on the scalar field, assuming that the gravitational fields are weak and the corresponding solutions are relativistic.

We will use a smooth potential $ V (s ^ 2) $, which has a minimum for a nonzero value of the argument $ V (s_1 ^ 2) = V_ {min} $, $ s_1 ^ 2> 0 $. For this minimum, the constant function $ \phi = s_1 $ is the exact solution to the problem. For such a function, using a spherical coordinate system and a metric of signature $ (- +++) $, we write out the mixed components of the energy-momentum tensor: 
\begin{eqnarray}
&&T_\mu^\nu=\diag(-\rho,p_r,p_t,p_t)=-V_{min}\cdot\diag(1,1,1,1),\\
&&\rho=V_{min},\ p_r=p_t=-V_{min},\\
&&\rho_{grav}=\rho+p_r+2p_t=-2V_{min}.
\end{eqnarray}
The result coincides with the standard EOS of dark energy, which explains the interest to this model in the cosmological context. We will fix $ V_ {min}> 0 $, and for simplicity we will assume $ V> 0 $ everywhere.

In this paper, we consider stationary spherically symmetric problems for which there are particular solutions of the form $ \phi = e ^ {iEt} s (r) $, with real $ E, s (r) $. With this substitution, the dimension is reduced $ (E ^ 2 + \Delta) s = 2V '(s ^ 2) s $. Next, we will consider stationary solutions $ E = 0 $, $ \phi = s (r) $. The uniqueness of solutions with stationary boundary conditions is demonstrated in the Appendix. Thus, all solutions that can be attached to the constant $ \phi = s_1 $ are globally stationary and have the form $ \phi = s (r) $.

Calculating EOS for stationary solutions 
\begin{eqnarray}
&&T_\mu^\nu=\diag(0,s'^2,0,0)-\diag(1,1,1,1)\cdot(s'^2/2+V(s^2)),\\
&&\rho=-p_t=s'^2/2+V(s^2)>0,\ p_r=s'^2/2-V(s^2),\\
&&\rho_{grav}=\rho+p_r+2p_t=-2V(s^2).
\end{eqnarray}
If the potential is shallow, then $ \rho_ {grav} \sim-2V_ {min} $, as for DE. This result is quite remarkable. As a consequence, the scenario can be configured in such a way that the gravitating density profile immediately after the NRDM phase $ \rho_ {grav} = \epsilon / (4 \pi r ^ 2)> 0 $ drops sharply to the DE phase $ \rho_ {grav} \sim-2V_ {min} $. This reproduces the phenomenological RDMcut scenario discussed in \cite {bled2020}, with a sharp cutoff of the density to almost zero at the $ R_ {cut} $ radius. The DE contribution begins to be felt at much larger distances and reproduces the observed effect of accelerated cosmological expansion there.

Technically, the condition of connection for the radial pressure component at the boundary between the phases must still be met. This condition can be satisfied if the model has enough degrees of freedom to ensure that in $ p_r $, the first term $ s'^2/2 $ dominates over the second $ -V (s ^ 2) $. In this case, it is possible to ensure the continuous connection with the positive $ p_r $ from the NRDM phase, no matter how large this value may be. Physical manifestations are defined only by $ \rho_ {grav} $ and do not depend on the details of this connection.

We will make such a connection for a particular choice of the potential. First of all, we write the right-hand side of the equations of motion in the form $ 2V '(s ^ 2) s = V (s ^ 2)' _ s $. Next, using the reparametrization of the argument $ V (s ^ 2) = V_1 (s) $, we choose the potential as given below. The remarkable properties of such a potential are the linearity of the equation of motion, the existence of an analytical solution, and also the fact that any potential in the vicinity of the minimum can be written as follows: 
\begin{eqnarray}
&&V_1(s)=V_{min}+a/2\,(s-s_1)^2,\ a>0,\ s_1>0,\label{V1q}\\
&&s''+2s'/r = a(s-s1),\\
&&s=s_1 + (e^{-\sqrt{a} r} C_1)/r + (e^{\sqrt{a} r} C_2)/(2 \sqrt{a} r)\label{V1sol}.
\end{eqnarray}
Selecting a branch with finite $ s \to s_1 $ at $ r \to \infty $, we get $ C_2 = 0 $. We also impose the condition $ C_1> 0 $ in order to ensure $ s> s_1 $ on the solutions. For $ s> s_1 $, the ascending branch of $ V_1 (s) $ corresponds to the positive square of the mass, normal particles. At that time, for $ s <s_1 $, the descending branch of $ V_1 (s) $ formally corresponds to the negative square of the mass, the tachyon case, but this branch is not used in the solutions we have considered. Calculating the components
\begin{eqnarray}
&&p_r=e^{-2 \sqrt{a} r} C_1^2 (1 + 2 \sqrt{a} r)/(2r^4) -V_{min},\\
&&\rho_{grav}=-a C_1^2 e^{-2 \sqrt{a} r}/r^2- 2V_{min},
\end{eqnarray}
we see that by choosing $ C_1 $ it is always possible to achieve a connection with positive $ p_r $ from the NRDM phase. At the same time, choosing small $ a $, one can reach $ \rho_ {grav} \sim -2V_ {min} $. With such a choice of parameters, the solution comes arbitrarily close to the RDMcut+DE profile shown in Fig.\ref {f2}, thereby providing a deeper physical foundation for it.

As before, exact matching with cosmological estimates can be achieved by comparing $ N '_ {gal} (M_ {dm, gal} + M_ {vac}) $, $ M_ {dm, gal} = \epsilon R_ {cut} c ^ 2 / G $, $ M_ {vac} = (8 \pi / 3) \rho_ {de, bgr} R_ {cut} ^ 3 $, with known $ M_ {dm, uni} \sim4.5 \cdot10 ^ {23} M_ \odot $. Exact matching is ensured, in particular, when choosing $ R_ {cut} = 50 $kpc, $ N '_ {gal} = 1.7 \cdot10 ^ {12} $, or $ R_ {cut} = 44 $kpc, $ N' _ {gal} = 2 \cdot10 ^ {12} $, or $ R_ {cut} = 0.6 $Mpc, $ N '_ {gal} = 1.4 \cdot10 ^ {11} $. 

\section {Addition: photon case} \label {sec3}

Since the photons of the Standard Model are not sterile, corrections are required to use them in the  described scenarios. Specifically, an analysis of three questions is required:

\begin {itemize}
\item generation of longwave photons by compact massive objects;
\item the passage of such photons through the interstellar medium;
\item Bose-Einstein condensation of photons.
\end {itemize}

In this article, we will only consider in detail the question of photon generation. The main difference from the sterile case is the interaction of photons at high energies, leading to the production of $ e ^ + e ^ - $ pairs and other particles. We will assume that these particles are localized in the ultrarelativistic plasma layer between the NRDM phase and the Planck core. Fortunately, the EOS of ultrarelativistic plasma is independent of its actual composition and even its temperature. Such a plasma is described by the universal TOV equation with a factor $ w = 1/3 $, as if the plasma consisted entirely of radiation. In this case, it is only important that the kinetic energies of plasma particles significantly exceed their rest masses, and also that the EOS is isotropic and has equal components of radial and transverse pressure.

The second question, about the possible passage of photons through the ISM, imposes a limitation on their frequency. Electromagnetic waves can propagate in the ISM only if their frequency exceeds Langmuir's value, which varies from a few kHz in the central regions of the galaxy to some Hz in the outer regions. On the other hand, if the wavelength becomes comparable to the size of galactic structures, then, presumably, waves can penetrate them without absorption, similar to long radio waves penetrating the walls of buildings and other structures. That is, it can be expected that the ISM transparency window, which closes at Langmuir's frequency, reopens at ultra-low frequencies.

Finally, the third question, about the possibility of Bose-Einstein condensation of photons, has been intensively discussed recently. In a complete vacuum, photons cannot condense, because they are massless, and the state of minimum energy for them coincides with the vacuum. At the same time, in \cite {1305.1210} and references therein it was noted that in ISM/IGM photons have a dispersion relation equivalent to the presence of a nonzero mass of a photon. As a result, there is a theoretical possibility that the photons in medium can undergo Bose-Einstein condensation.

Now we will consider the question of photon generation by the NRDM|TOV system. The required equations are listed in the Appendix. The equations are formulated for the metric profiles $ A $ and $ B $, in the logarithmic representation: $ x = \log r $, $ a = \log A $, $ b = \log B $. Hereinafter, $ A = -g_ {tt} $ and $ B = g_ {rr} $ denote the temporal and radial components of the metric tensor, which completely describe the structure of the gravitational field for stationary spherically symmetric problems. 

\begin{figure}
\begin{center}
\includegraphics[width=\textwidth]{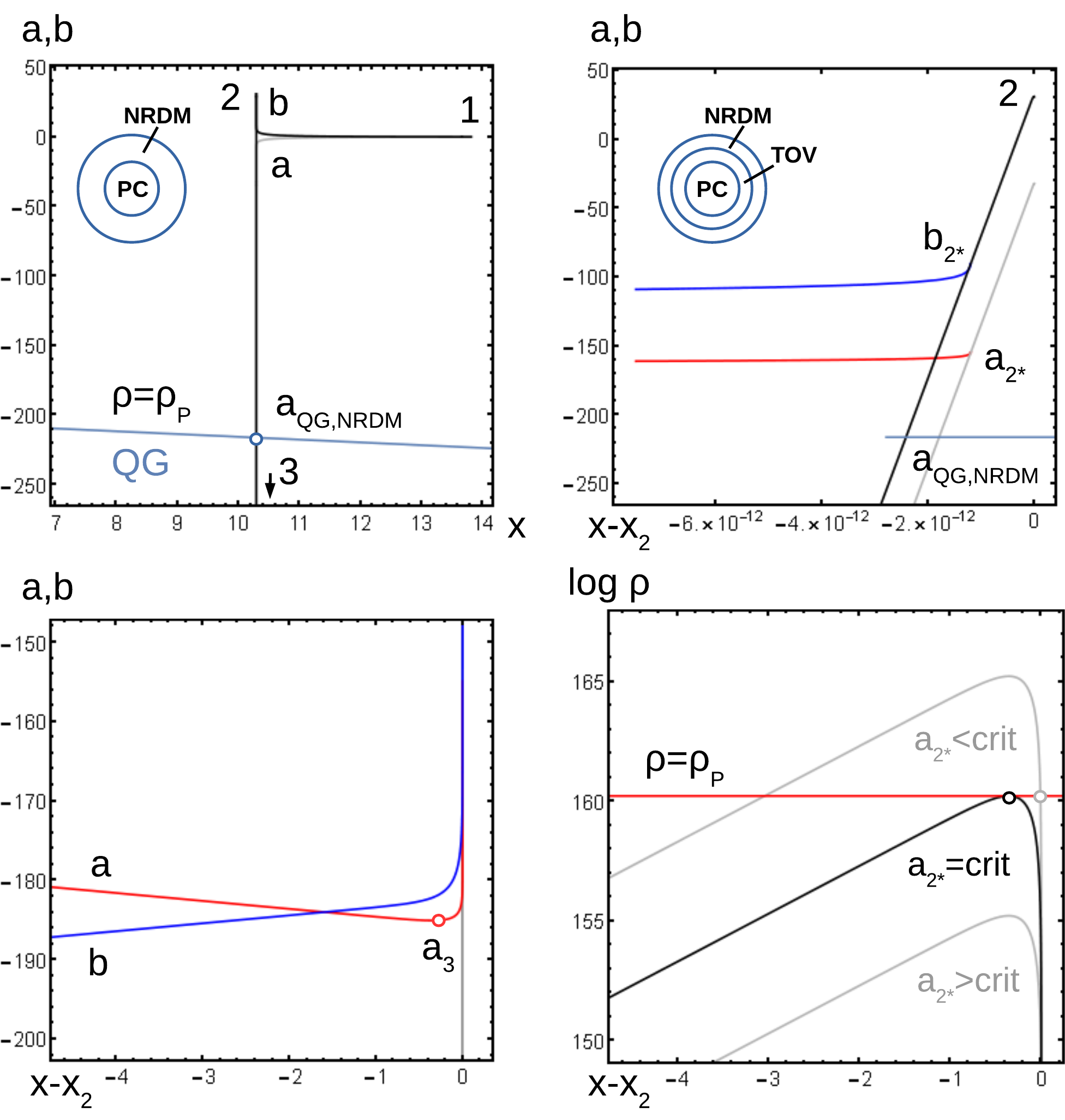}
\end{center}
\caption{Solutions with NRDM, TOV and PC (Planck core) phases. See text for details.}\label{f4}
\end{figure}

\begin{table}
\begin{center}{\footnotesize
\caption{NRDM|TOV-star, stellar mass, critical case}\label{tab1}

~

\def\arraystretch{1.1}
\begin{tabular}{|c|c|}
\hline
model parameters&$M=10 M_\odot$, $r_{s}=29532.4$m, $\epsilon=10^{-14}$\\ \hline
starting point&$a_1=0$, $b_1=0.0299773$, \\ 
of integration& $r_1=10^6$m, $L_1=1.05738\cdot10^6$m\\ \hline
supershift&$a_{2}=-32.2362$, $b_{2}=30.8799$, \\ 
begins&$r_{2}/r_s-1=5.26127\cdot10^{-7}$, $L_2=5.819\cdot10^{-3}$m\\ \hline
NRDM|TOV&$a_{2*}=-154.936$, $b_{2*}=-90.434$, \\ 
transition&$r_{2*}/r_2-1=-1.2\cdot10^{-12}$, $L_{2*}=1.32074\cdot10^{-29}$m\\ \hline
supershift&$a_{3}=-185.052$, $b_{3}=-182.216$, \\ 
ends&$r_{3}=20638$m, $L_3=3.01835\cdot10^{-36}$m\\ \hline
minimal radius,& $a_4=-95.1895$, $b_4=-272.808$, \\ 
end of integration&$r_4=l_P=1.62\cdot10^{-35}$m, $L_4=0$\\ \hline
\end{tabular}

}\end{center}
\end{table}

\begin{table}
\begin{center}{\footnotesize
\caption{various scenarios, global parameters}\label{tab2}

~

\def\arraystretch{1.1}
\begin{tabular}{|c|c|}
\hline
\multicolumn{2}{|c|}{NRDM, $E_{in}=E_P$, $\lambda_{in}=l_P$} \\ \hline
$M=10M_\odot$, $\epsilon=10^{-12}$ & $A_{QG}=1.2\cdot10^{-92}$, $\lambda_{out}=4.8\cdot10^{-6}$pc\\
$M=10M_\odot$, $\epsilon=10^{-16}$ & $A_{QG}=1.2\cdot10^{-96}$, $\lambda_{out}=4.8\cdot10^{-4}$pc\\
$M=4.06\cdot10^6M_\odot$, $\epsilon=10^{-7}$ & $A_{QG}=7.2\cdot10^{-99}$, $\lambda_{out}=6.2\cdot10^{-3}$pc\\
$M=4.06\cdot10^6M_\odot$, $\epsilon=10^{-10}$ & $A_{QG}=7.2\cdot10^{-102}$, $\lambda_{out}=0.2$pc\\
\hline
\multicolumn{2}{|c|}{NRDM|TOV critical, $E_{in}=0.512$MeV, $\lambda_{in}=2.42\cdot10^{-12}$m} \\ \hline
$M=10M_\odot$, $\epsilon=10^{-12}$ & $A_{2*}=5.15\cdot10^{-70}$, $\lambda_{out}=3.46$Mpc\\
$M=10M_\odot$, $\epsilon=10^{-14}$ & $A_{2*}=5.15\cdot10^{-68}$, $\lambda_{out}=0.346$Mpc\\
$M=10M_\odot$, $\epsilon=10^{-16}$ & $A_{2*}=5.15\cdot10^{-66}$, $\lambda_{out}=0.0346$Mpc\\
$M=4.06\cdot10^6M_\odot$, $\epsilon=10^{-7}$ & $A_{2*}=2.97\cdot10^{-86}$, $\lambda_{out}=4.56\cdot10^5$Gpc$\gg R_{uni}$\\
$M=4.06\cdot10^6M_\odot$, $\epsilon=10^{-10}$ & $A_{2*}=2.94\cdot10^{-83}$, $\lambda_{out}=1.45\cdot10^4$Gpc$\gg R_{uni}$\\
\hline
\end{tabular}

}\end{center}
\end{table}

We numerically solve these equations using {\it Mathematica} {\tt NDSolve} algorithm. As noted in \cite {bled2020}, TOV systems are characterized by critical phenomena, abrupt changes in the solution with continuous change in parameters. As a result, the NRDM|TOV system has a richer solution structure than a pure NRDM.

Fig.\ref {f4} shows the solution for a compact object of stellar mass $ M = 10M_ \odot $. At the same time, as the study of the rotation curves \cite {bled2020} shows, the parameter $ \epsilon $ for the external NRDM phase can be chosen in the interval $ 10 ^ {- 12 ...- 16} $, and here we choose it in the center of this interval : $ \epsilon = 10 ^ {- 14} $.

The top left image shows the behavior of a pure NRDM solution, similar to the graphs from \cite {bled2020} for a supermassive solution. The solution starts at point 1, located at a large distance from the object, in the weak field region. Then, near the gravitational radius, the solution tries to enter the Schwarzschild mode with symmetrically diverging $ a $ and $ b $ profiles. The $ b $ profile reaches its maximum at point 2, which is very close to the Schwarzschild radius $ r_s $. After passing point 2, the solution goes into the supershift or mass inflation mode \cite {gr-qc/0411062}. In a thin layer near this point, the $ a $ and $ b $ profiles rapidly decrease with decreasing $ x $, while the mass density rapidly increases. At the point $ a_ {QG} $, the mass density reaches the Planck value. At this point, the NRDM phase joins the Planck core (PC). We recall that in the models under consideration, the Planck core has a large effectively negative mass due to quantum effects. This mass produces a force of gravitational repulsion, which maintains in equilibrium a coat of large positive mass located in the supershift region. These two masses almost cancel each other out, so that the object as a whole has an initial stellar mass.

The top right image shows the modification of the solution when it is connected with the TOV phase. The $ a $ and $ b $ profiles rapidly falling in a thin $ x $-layer of the NRDM phase, before passing the Planck boundary and after crossing the $ 2 * $ phase boundary, are replaced by more slowly falling TOV phase profiles. In the figure at the bottom left, the TOV solution is continued into a wider $ x $-layer. At point 3, the solution passes through the minimum of the $ a $ profile. Further, the profiles $ a $ and $ b $ diverge symmetrically in the region of large negative values, which is typical for the Schwarzschild singularity of negative mass \cite {Blau} Sec.28.5. Unlike other known solutions, this singularity is not naked, it is covered with a massive coat and remains in the superstrong redshift zone until the integration stops at the Planck radius.

The figure at the bottom right shows the behavior of the mass density of the solution in the TOV phase. The maximum density is reached at the minimum point $ a_3 $. The {\it critical solution}, which touches the Planck density line, is specially highlighted. For this solution, all the other TOV graphs in this figure are shown. The key parameters for this solution are also listed in Table~\ref {tab1}.

Upon reaching the Planck density, the solution joins the Planck core. {\it In supercritical mode}, that is, when the point of joining of the phases $ a_ {2 *} $ is selected below the critical position, the density reaches the Planck value earlier, the boundary of the Planck core shifts accordingly.

{\it In subcritical mode}, the Planck density is not reached at all, and the solution continues until the central singularity. This singularity is quite similar to the Planck core, since it also has a large negative mass, and its repulsive force keeps the system in equilibrium. The singularity can be smeared over the $ r \sim l_P $ neighborhood to obtain a regular core. One can also expand the core to larger radius values. We will call such solutions the Planck core of type II, in contrast to the previously considered solutions, to which we assign the type I. Their difference is that core I arises when the NRDM or TOV matter reaches the Planck density, while for core II the density of this matter remains sub-Planckian, and the core consists of other matter exceeding the Planck density. NRDM solutions for the physically relevant selection of parameters necessarily exceed the Planck density, so a type I core is formed there. For the TOV phase, both types of solutions are possible.

\paragraph* {Wavelengths} of photons in the resulting gravitational field are easy to calculate. The initial wavelength, which was equal to $ \lambda_ {in} = l_P $ on the surface of the Planck core for the NRDM solution, is replaced by $ E_ {in} = 0.512 $MeV, $ \lambda_ {in} = 2.42 \cdot10 ^ {- 12 } $m at the interface for the NRDM|TOV solution. This choice corresponds to the threshold for the production of $ e ^ + e ^ - $ pairs in the collision of incoming and outgoing photon fluxes. An ultrarelativistic TOV plasma is composed of these pairs and other higher-energy particles. Applying the redshift to this wavelength for the critical case $A_{2*}=5.15\cdot10^{-68}$ from Table~\ref {tab1}, we get $ \lambda_ {out} = \lambda_ {in} / \sqrt {A_ {2 *}} = 1.07 \cdot10 ^ {22} $m $ = 0.346 $Mpc. In the supercritical mode, longer wavelengths are obtained, in the subcritical mode, shorter ones. Note the coincidence of the obtained wavelength with the characteristic size of the MW galaxy. This coincidence is even more surprising if we note that the metric coefficients appearing in the intermediate calculations vary in the considered solutions by a hundred orders of magnitude.

\paragraph* {L-integral.} Table~\ref {tab1} also lists the values of the invariant length integral measured in the radial direction: $ L = \int dr \sqrt {B} $. Integration starts from the minimum radius $ r = l_P $. The integral up to point 3 turns out to be less than the Planck length, which does not pose a problem, since this region contains a Planck core with unknown properties, and the integration over this region should not be carried out at all. The integral up to point $ 2 * $ represents the thickness of the TOV layer. It is noteworthy that in the $ r $-coordinate this thickness is about 9km, while in the $ L $-coordinate this thickness is microscopic: $ L_ {2 *} = 8 \cdot10 ^ 5l_P $. This difference is an indicator of a strong deformation of the radial direction in the solution, which manifests itself in an extremely small $ B $-factor. Note that the ``aerial'' $ r $-coordinate is related to the area of the corresponding sphere (always equal to $ 4 \pi r ^ 2 $), while the $ L $-coordinate is the more appropriate local thickness characteristic. The resulting thin layer corresponds to a superdense TOV plasma sandwiched between the Planck core and NRDM coat under tremendous pressure. We remind that in our calculations we consider all the media to be continuous, disregarding their microscopic structure. Apparently, a quantum calculation should be carried out in this region, after which the results obtained here should be revised. In this work, we restrict ourselves to classical calculations. Further, up to point 2, the coordinate $ r $ changes microscopically by $ \sim3.5 \cdot10 ^ {- 8} $m, while the change in $ L $ is about 6mm. In this region, the deformation becomes inverse, due to the large $ B $-factor near point 2. Finally, in the outer region, an $ L $-integral is accumulated, comparable to the change in $ r $, since the region of weak fields prevails here, and the space becomes flat.

The picture obtained in the analysis of the $ L $-integral coincides with the structure of compact massive objects and the presence of a ``shrinking volume'' \cite {1612.04889} inside them. In the region shallow in the $ L $-coordinate, thin superdense layers of matter are localized, creating strong deformations of space, strong gravitational fields that determine the structure of the solution as a whole.

\paragraph* {Sensitivity to model parameters.} Table~\ref {tab2} shows the calculation results for different scenarios. For a pure NRDM model, the value of $ \lambda_ {out} $ varies in the range of $ 10 ^ {11 ... 16} $m or $ 10 ^ {- 6 ...- 1} $pc. If the DM particles in this scenario were photons, their propagation in the ISM would be prohibited, since their frequency is significantly less than Langmuir's value and the wavelength is significantly less than the size of the galaxy. However, this scenario is limited to the sterile case and does not consider photons as DM particles. The photon case assumes an NRDM|TOV combination, and here, for compact objects of stellar mass, wavelengths are obtained in the range of 34.6kpc-3.46Mpc, that is, comparable to galactic sizes. We assume that such waves can propagate in ISM and condense in BEC, thereby providing a logical closure for the considered model. For the supermassive case, typical for central black holes in galaxies, $ \lambda_ {out} $ values are obtained that are much larger than the size of the universe $ R_ {uni} \sim14 $Gpc. This means that either supermassive black holes in this scenario do not participate in the formation of dark matter, or cosmological processes must be taken into account for their analysis. Note also that in the case of a pure NRDM for the described parameters there are exact formulas: $ A_ {QG} = \epsilon (l_P / r_s) ^ 2 / (8 \pi) $, $ \lambda_ {out} = r_s (8 \pi / \epsilon) ^ {1/2} $. For the NRDM|TOV combination, we currently have only empirical relationships. Namely, for the data in Table~\ref {tab2}, with fixed $ r_s $ and variable $ \epsilon $, the relations hold $ A_ {2 *} \sim \epsilon ^ {- 1} $, $ B_ {2 *} \sim \epsilon ^ {- 3} $, $ \lambda_ {out} \sim \epsilon ^ {1/2} $, which are the sign of hidden symmetries in the considered system.

\paragraph* {Qualitative analysis.} At the first glance, the presented results are relevant only to the model of NRDM|TOV stars studied here. Now we will show that many features of the described solution are typical for a wider class of models, possibly for all compact massive objects \cite {0902.0346,1612.04889,1401.6562,1409.1501}. First of all, {\it erasure of the event horizon} is associated with the T-symmetry of solutions, which results in a grid of the intersecting flows of incoming and outgoing particles. For true black holes, there are only incoming flows, and for T-conjugated white holes, only outgoing flows. In the solutions under consideration, there can be no horizons that could prevent the entry or exit of the particles. Another explanation for the horizon erasure effect arises when observing the mass function $ M (r) $. The region below the horizon corresponds to $ 2M (r) / r> 1 $. In the presence of distributed positive mass density, the function $ M (r) $ decreases with decreasing $ r $ (one can imagine how positive mass layers are removed from the solution). If the density is high enough, then $ 2M (r) $ decreases faster than $ r $, so no horizon is formed. High density arises from the phenomenon of {\it mass inflation} \cite {gr-qc/0411062}. This phenomenon occurs due to the positive feedback between pressure and gravity. In strong gravitational fields, in equilibrium systems, the pressure increases rapidly with decreasing radius, due to the hydrostatic equation. At high pressure, it begins to contribute to gravity on the same basis as mass density, which also increases due to EOS. As a result, the gravitational field is strengthened, which leads to a further increase in pressure and density. As a result, a thin layer is formed in the solution, in which pressure, density and gravitational field increase very rapidly. Further, the rapid decrease in the mass function $ M (r) $ does not stop at zero crossing, {\it the mass becomes negative}. While the local density is still positive, the value of the central mass located under a certain radius $ r $ is negative. A successful closure of the model is the concept of the {\it Planck core}, according to which, when the Planck density is exceeded, the quantum corrections make the mass effectively negative \cite {1401.6562,1409.1501}. We described another possibility in this work, when a central singularity or a superdense core, consisting of matter other than the surrounding massive coat, has a negative mass. Next, we observe the behavior of the {\it metric profiles} $ A $ and $ B $, where the first describes the time dilation and redshift, and the second describes the deformation of the radial coordinate. The mass function is directly related to the $ B $-profile by the formula $ B = (1-2M / r) ^ {- 1} $. Large negative masses correspond to a small positive $ B $. As a result, the integral of length $ L = \int dr \sqrt {B} $ in this region becomes small, the region becomes {\it shallow} in $ L $, although in $ r $ it can occupy an essential part of the solution. Together with the coefficient $ B $, $ A $ also becomes small in the region of mass inflation, which is typical for strong gravitational fields. This leads to {\it a strong time dilation and redshift}, as a result, from the point of view of an external observer, the object is dark, almost like a black hole. At the same time, high energies inside the object can result in high density radiation. The redshift does not affect the flux density in the transverse direction and shifts the radiation to the low-energy region. In the models under consideration, the superstrong redshift stretches photons into ultra-long wave packets that can reach the size of a galaxy, while a large number of such photons, after taking into account all sources, leads to a significant contribution to the mass of the galaxy. Thus, microscopically thin high-energetic layers inside compact objects appear to be conjugated with galactic-scale structures. At large distances from the center, the radiation density decreases as $ \rho \sim r ^ {- 2} $, which leads to flat rotation curves of galaxies. The distribution of black holes in the galaxy modulates these dependencies and allows to describe {\it the observed rotation curves} with their deviations from the flat shape \cite {bled2020}. These calculations do not depend on the nature of the emitting objects, only on the assumption that their distribution is proportional to the luminous matter. Thus, radiation from compact massive objects, be they photons or other particles, may be directly unobservable due to long wavelengths. However, it can determine the rotation curves of galaxies and produce other gravitational effects that are usually associated with {\it dark matter}. If this type of radiation can pass the interstellar medium and become Bose-Einstein condensate outside the galaxy, one will simultaneously obtain the description of {\it dark energy}. 

\section{Conclusion}

This work continued the construction of a recently proposed model of NRDM-stars. In this model, the quanta dark of matter are sterile massless particles that are emitted from quasi-black holes located mainly in the central regions of the galaxy. At large distances from the center of the galaxy, the emission is directed radially and the mass density has the form $ \rho \sim r ^ {- 2} $, in contrast to the homogeneous model $ \rho = Const $. In the cosmological context, the homogeneous model with massless particles corresponds to the radiation epoch of the expansion of the universe, while the proposed inhomogeneous model turns out to be equivalent to $ \Lambda $CDM.

Specifically, several scenarios were considered in which the radial emission of dark matter is brought into hydrostatic equilibrium with a uniform background:

\begin {itemize}
\item [S1.1:] a jump in the dark energy density at the edge of the galactic halo;
\item [S1.2:] a surface tension at the boundary between the halo and the dark energy background;
\item [S1.3:] a phase transition of dark matter into dark energy, accompanied by the Joule-Thomson effect;
\item [S1.4:] Bose-Einstein condensation of dark matter inside the galaxy into dark energy outside of it.
\end {itemize}

From the junction conditions, the density correspondence to the observed $ \Omega $-parameters and mass functions typical for $ \Lambda $CDM model were obtained. In these scenarios, CDM macro-particles are galaxies with massive halos surrounding them, floating in a homogeneous medium with the dark energy equation of state.

Additionally, the question of what happens if dark matter particles are not sterile, for example, are photons of the Standard Model, is considered. When high-energy photons collide inside NRDM stars, as a result of the production of $ e ^ + e ^ - $ pairs and other particles, a thin layer of ultrarelativistic plasma appears. The performed classical calculation shows that the photons emitted from this layer, after applying the gravitational redshift, acquire a wavelength comparable to the galactic sizes. For the logical closure of the photon scenario in this model, it is also necessary to study the questions of the passage of such longwave photons through the interstellar medium and Bose-Einstein condensation of the photons in the intergalactic medium. The study of these questions will be continued. 

\section*{Acknowledgements}
The author thanks the organizers and participants of the Bled 2021 Workshop ``What comes beyond the Standard models?'' for fruitful discussions. The author also thanks Kira Konich for proofreading the paper.

\footnotesize

\subsection*{Appendix: Details of constructions}

\paragraph {Modeling the outer part of the rotation curve.} Fig.\ref {f2} on the left shows the outer part of the MW rotation curve, overlaid with various model profiles. RDMcut represents the simplest radial dark matter halo model introduced in \cite {bled2020}, which is cut off at the $ R_ {cut} $ radius, so that the velocity on the outer portion of the curve is $ v = (GM_ {dm} (R_ {cut} ) / r) ^ {1/2} $. The graph also shows the standard galactic profiles of Navarro-Frenk-White (NFW) and Einasto, with parameters adjusted to the experimental points. The source of experimental data is the work \cite {1307.8241}. It can be seen that all the profiles pass approximately the same way in the corridor of experimental scatter. Since the scatter in the outer region of the rotation curve is extremely large, it is not possible to select any particular profile based on this data.

Fig.\ref {f3} shows the profiles obtained in the accepted scenarios of this work, compared with the RDMcut profile, with the parameters $ \epsilon = 2.5 \cdot10 ^ {- 7} $, $ R_ {cut} = 50 $kpc. The profiles are related to the gravitating mass by the relation $ v = (GM_ {grav} (r) / r) ^ {1/2} $. The characteristic downward bend of all new profiles at large distances corresponds to the negative contribution of dark energy growing under the root. With a large value of the radius, the gravitating mass becomes negative, and the corresponding radial acceleration also changes sign. In this zone, $ v $ is not defined, circular orbital motion is impossible, here the accelerated cosmological expansion of the universe begins to dominate.

All new profiles pass close to RDMcut and with it fit into the corridor of errors. The S1.4 profile is effectively the same as RDMcut if DE contribution is included in it. Profile S1.3 for the parameters selected above is shown in Fig.\ref {f3} right in blue. It can be pulled closer to RDMcut by setting $ R_ {cut1} = 30 $kpc, as a result, the key parameters of the model will slightly change: $ \epsilon = 2.5 \cdot10 ^ {- 7} $, $ \{R_ {cut1}, R_ { cut1b}, R_ {cut2} \} = \{0.03,0.20,0.55 \} $Mpc, $ \{M_1, \Delta M_1, \Delta M_2, M_ {vac} \} = \{1.56, 2.07, - 1.56, 1.41 \} \cdot10 ^ {11} M_ \odot $, $ M_ {dm + de, gal} + M_ {vac} = 3.5 \cdot10 ^ {11} M_ \odot $, while the cosmological estimates are performed with new values $ \rho_1 / \rho_2 = 2288 $, $ N '_ {gal} = 1.3 \cdot10 ^ {12} $.

Let's pay attention again to Fig.\ref {f2} left. Noteworthy is the presence of a rise in the experimental curve at the exterior of this graph. It is responsible for the Local Group (LG) structures outside of the MW. Recall that the galaxy M31/Andromeda closest to the MW is located at a distance of $ r_ {M31} \sim0.8 $Mpc. Therefore, at the exterior of the curve, the radius begins to capture M31 and other LG structures, increasing the total gravitating mass. Note that spherical symmetry is lost in this case, and the formula $ v = (GM_ {grav} (r) / r) ^ {1/2} $ should no longer be used for the velocity, except as a rough approximation. In the fit \cite {bled2020}, shown in green on the graph, the RDMcut contribution of dark matter was taken into account, as well as the contribution of luminous matter, which is active at small radii, and the region increasing at large LG radii was described empirically as an additive homogeneous background. The density value obtained at the fit was $ \rho_ {bgr} \sim4.4 \cdot10 ^ {- 26} kg / m ^ 3 $, which is 4.4 times higher than the critical cosmological density. In this interpretation, this contribution has nothing to do with the cosmological background, it only describes the overdensity averaged by the Local Group, limited in space.

For a more detailed analysis, Fig.\ref {f2} right shows the raw data from \cite {0811.0860}, on the basis of which the experimental rotation curve \cite {1307.8241} was built. Strictly speaking, Fig.\ref {f2} right shows not the rotation curve, but the dependence of the experimentally measured radial velocity component on the distance. The outer part of this dependence is taken from \cite {0811.0860}, Table 1, SF sample, $ r_ {GC} $ and $ v_ {GC} $ columns. The radial velocities Fig.\ref {f2} right have a sign, in contrast to Fig.\ref {f2} left, where the absolute value of the velocity is given. The pattern is striking: on the outer part of the curve $ v_r> 0 $, which corresponds to expansion, on the inner part of $ v_r <0 $, there is a contraction. The outer part can be approximated by the Hubble law $ v = Hr $ with the value $ H = 95 \pm16 $km/s/Mpc. Compared to recent cosmological estimates $ H_0 = 68-77 $km/s/Mpc, the average value is somewhat overestimated, but fits within 1.7 standard deviations. Thus, the outer part of the curve is consistent with the Hubble flow. It was noted in \cite {astro-ph/9711073} that Hubble's law begins to operate directly outside of LG, here we see that it also operates on the outer border of LG. The inner part of the graph Fig.\ref {f2} on the right corresponds to the collapse of matter under the influence of gravitation towards massive galaxies that are part of LG. Recall that LG consists of two large galaxies, MW and M31, and many globular clusters and satellite galaxies. For these satellites, the radial velocities were measured relative to the MW, which are used to construct the outer part of the rotation curve \cite {1307.8241}.

In the work \cite {1405.0306}, a model of the outer region of LG is described, which reproduces just such a picture, Hubble flow at the outer boundary of the region and the collapse of matter to the center of mass of LG at the inner boundary. In \cite {1405.0306}, Fig.10 qualitatively coincides with our Fig.\ref {f2} right. The details show differences caused by using a different coordinate system and a different dataset for analysis. The work \cite {1405.0306} also gives estimates of the total masses of MW and M31: $ M_ {MW} = (0.8 ^ {+ 0.4} _ {- 0.3}) \cdot10 ^ {12} M_ \odot $ and $ M_ {M31} = (1.5 ^ {+ 0.5} _ {- 0.4}) \cdot10 ^ {12} M_ \odot $. Since \cite {1405.0306} uses an additive homogeneous contribution of dark energy, for comparison with our scenario, we need to take into account the $ M_ {vac} $ term, which in the above scenarios gives the value $ M_ {MW} = (0.3-0.5) \cdot10 ^ {12} M_ \odot $, within 1-1.7 standard deviations from the value \cite {1405.0306}. As for Andromeda, \cite {0811.0860} assumed that the masses of MW and M31 were equal, and LG required additional mass that is not part of these galaxies. Whereas in \cite {1405.0306} scenarios with M31 mass 2 times the MW mass are preferred, and additional LG mass is not required.

In our work, we concentrate on describing internal orbits closely bound to the MW, for which the masses of external structures are not important. The exact solution to the spherically asymmetric problem of the interaction of nearby galaxies with overlapping halos is rather nontrivial. The resulting halo shape can deviate from the sphere. The gravitational field is not described by a simple rotation curve depending only on the radius. The result depends on a variety of model assumptions such as dark matter EOS and initial conditions. Therefore, in this work, we prefer to restrict ourselves to cosmological estimates for spherically symmetric halos of separately standing galaxies and will not consider the cases of nearby galaxies overlapping by the outer parts of their halos. In practice, this means that in the MW rotation curve we will consider only the descending part, cutting off its LG tail. 

\paragraph * {Model independent reconstruction of EOS.} Let us describe another algorithm that, for spherically symmetric halos, in principle, can reconstruct EOS directly from the rotation curve, without the assumptions about the phase transition path made in scenario S1.3. Let us know the rotation curve for an average-mass galaxy, approximated by some empirical profile. The mass function $ M_ {grav} (r) $ and the density $ \rho_ {grav} (r) $ are trivially recovered from it. On the outside of the rotation curve, these two functions are matched with cosmological estimates: $ M_ {grav} (R_ {max}) + M_ {vac} (R_ {max}) = M_ {dm, uni} / N_ {gal} $, $ M_ {vac} (r) = (8 \pi / 3) \rho_ {de, bgr} r ^ 3 $ and $ \rho_ {grav} (R_ {max}) = - 2 \rho_ {de, bgr} $. Note that these conditions are imposed directly on the experimental curves and not on the EOS components. We use the relations $ \rho_ {grav} = \rho + p_r + 2p_t $ and $ r (p_r) '_ r + 2p_r-2p_t = 0 $, these are two relations for three profiles $ (\rho, p_r, p_t) $. As a result, one functional degree of freedom remains. One can set an arbitrary function $ p_r $, then $ (\rho, p_t) $ will be reconstructed by linear formulas, even without solving differential equations. Boundary conditions on the outer radius are of the form $ p_r = p_t = - \rho $, in this case, due to the conditions imposed above, it will automatically satisfy $ \rho = \rho_ {de, bgr} $, $ p_r = p_t = - \rho_ {de, bgr } $, $ p_ {r} '= 0 $. Using these constraints on $ (p_r, p_ {r} ') $ as boundary conditions, we can construct EOS in parametric form $ (\rho, p_r, p_t) (r) $. This construction can be supplemented with boundary conditions for NRDM at the inner radius $ \rho = p_r $, $ p_t = 0 $, by introducing the gravitational term into the hydrostatic equation, and other model corrections.

A similar algorithm for recovering EOS from rotation curves was used in the work \cite {1301.6785}. In this case, EOS was assumed to be isotropic $ p_r = p_t $, as a result, the solution did not contain functional ambiguities, but the NRDM-type solution was missed. The main obstacle to the implementation of such algorithms is the extremely large scatter in the outer region of the rotation curves, which allows different empirical profiles and leads to an inaccurate reconstruction of EOS in this region.

\paragraph {Taking into account the mass distribution of galaxies.} The above calculations use the estimated value of the number of galaxies $ N_ {gal} = 2 \cdot10 ^ {12} $ from \cite {1607.03909v2}. This value takes into account the evolution of the universe over time and represents an estimate of the number of observed galaxies up to redshift values $ z <8 $. In fact, to compare with the density of dark matter today, we need the number of galaxies in a simultaneous slice in a ball of radius $ R_ {uni} \sim14 $Gpc. This radius is purely nominal, the final formulas include the ratio $ M_ {dm, uni} / N_ {gal} $, from which this radius drops out. In fact, we need an estimate of the density of galaxies $ dN_ {gal} / dV $ near our position, for small $ z $. The mentioned ratio is expressed through this density: $ M_ {dm, uni} / N_ {gal} = \rho_ {dm} / (dN_ {gal} / dV) $.

In \cite {1607.03909v2}, the density of galaxies is modeled using the Schechter function:
\begin{equation}
dN_{gal}/dV/dM=\phi^*\log(10)10^{(M-M^*)(1+\alpha)}\exp(-10^{(M-M^*)}),
\end{equation}
where $ M = \log_ {10} (M_ {lm, gal} / M_ \odot) $, $ M_ {lm, gal} $ is the stellar mass of the galaxy. For the rest of the parameters, the values from the second row of Table~1 in \cite {1607.03909v2} were selected, representing the most accurate fit for the galaxies closest to us: $ \alpha = -1.29 $, $ M ^ * = 11.44 $, $ \phi ^ * = 12.2 \cdot10 ^ {- 4} Mpc ^ {- 3} $. Integrating this expression over the interval $ 6 \leq M \leq12 $ shown in Fig.1 in \cite {1607.03909v2}, we obtain $ dN_ {gal} /dV=0.154Mpc ^ {- 3} $, multiplying by $ (4 \pi / 3) R_ {uni} ^ 3 $, we get $ N_ {gal} = 1.766 \cdot10 ^ {12} $. It is noteworthy that the obtained value is close to the number $ 2 \cdot10 ^ {12} $, which was found in \cite {1607.03909v2} for the same mass interval and took into account the evolution of the universe.

Next, we need the mean $ <v ^ 2> $ for the square of the outer orbital velocity, on the same distribution. To find it, we use the Tully-Fisher relation $ v \sim (M_ {lm}) ^ p $ with exponent $ p = 1/4 $. Let us introduce a normalization to the value of MW and denote $ \eta_p = <(M_ {lm} / M_ {lm, MW}) ^ p> $, so that $ <v ^ 2> / v_ {MW} ^ 2 = \eta_ {1 / 2} $. Using the value $ M_ {lm, MW} = 6.08 \cdot10 ^ {10} M_ \odot $ from \cite {1407.1078} and calculating the average, we get $ \eta_ {1/2} = 0.0455 $, $ N_ {gal} \eta_ {1/2} = 1.196 \cdot10 ^ {11} $. This estimate is based only on experimental data in the form of Schechter and Tully-Fisher relations. It needs to be compared with the corrected $ N '_ {gal} $ parameter in our scenarios.

Before proceeding to the comparison, note that for the integration we have chosen the lower limit $ M_ {min} = 6 $, as in \cite {1607.03909v2}. This limit is slightly below the experimental data collection limit $ M_ {min} = 8 $, that is, extrapolation is used in the calculations. Note that the number of galaxies strongly depends on this limit. If we take $ M_ {min} = 8 $, we get $ N_ {gal} = 4.189 \cdot10 ^ {11} $. At the same time, the value of $ \eta_ {1/2} $ will increase by approximately the same factor and the value of $ N_ {gal} \eta_ {1/2} $ will practically not change. The same effect is observed for all $ p> 0.3 $. The reason for this is that the cumulative value of $ N_ {gal} \eta_p $ is expressed by an integral dominated by large masses.

Also note that the modeling width for Schechter function Fig.1 \cite {1607.03909v2} is 0.4-1dex, and the scatter width of Tully-Fisher relation \cite {astro-ph/0609076} for $ v ^ 2 $ is about 0.8dex. Therefore, deviations of $ <$ 1.8dex in comparison of model and experiment can be tolerated.

Most of our scenarios have a clear algebraic structure, producing an analytical answer of the form $ M_ {dm, uni} = N_ {gal} (M_ {dm, gal} + M_ {vac}) $, $ M_ {dm, gal} = k_1 \epsilon R_ {cut} c ^ 2 / G $, $ M_ {vac} = k_2 (8 \pi / 3) \rho_ {de, bgr} R_ {cut} ^ 3 $. The constants $ k_ {1,2} $ for scenarios $ \{S1.1, S1.2, S1.4 \} $ are of the form $ k_1 = \{2 / 3,3 / 4,1 \} $, $ k_2 = \{0,1 / 4,1 \} $. In calculations in order of magnitude, for fixed $ N_ {gal} = 2 \cdot10 ^ {12} $, $ M_ {dm, uni} \sim4.5 \cdot10 ^ {23} M_ \odot $, and for $ R_ { cut} <0.6 $Mpc, the contribution of $ M_ {vac} $ can be neglected. Also, in order of magnitude, we can consider $ k_1 \sim1 $. As a result, we get the only relation $ M_ {dm, uni} \sim N_ {gal} \epsilon R_ {cut} c ^ 2 / G $ for these three scenarios, which we must check with the experiment.

Next, we will make two estimates. In the first, for scenarios $ \{S1.1, S1.2, S1.4 \} $, we will assume that $ R_ {cut} $ is fixed, and $ \epsilon $ is distributed over galaxies. In this case, the relation has the form $ M_ {dm, uni} \sim N_ {gal} <\epsilon> R_ {cut} c ^ 2 / G $. The same relationship is obtained if we assume that $ R_ {cut} $ is distributed, but uncorrelated with $ \epsilon $, in this case it will be $ M_ {dm, uni} \sim N_ {gal} <\epsilon> <R_ {cut}> c ^ 2 / G $. Further, taking into account $ \epsilon = (v / c) ^ 2 $ and using the value $ \eta_ {1/2} $ introduced above, we get $ M_ {dm, uni} \sim N_ {gal} \eta_ {1/2 } M_ {dm, MW} $, $ M_ {dm, MW} = \epsilon_ {MW} R_ {cut} c ^ 2 / G $. It is convenient to rewrite this relation as $ N '_ {gal} \sim N_ {gal} \eta_ {1/2} $, where $ N' _ {gal} = M_ {dm, uni} / M_ {dm, MW} $ is the corrected number of galaxies introduced above in scenarios with MW copies. Substituting here $ \epsilon_ {MW} = 2.5 \cdot10 ^ {- 7} $, with $ R_ {cut} $ varying within 50kpc-0.6Mpc we get $ N '_ {gal} = 1.7 \cdot10 ^ {12} - 1.4 \cdot10 ^ {11} $, which coincides with the experimental estimate $ N_ {gal} \eta_ {1/2} = 1.196 \cdot10 ^ {11} $ within 1.2-0.1dex, with preference for large values of $ R_ {cut } $.

Next, let's make the estimation for the S1.3 scenario. In it, the adjustment of the $ \epsilon $ and $ R_ {cut} $ parameters is not as easy as in other scenarios, unless additional assumptions are made about the scaling of galaxies. As a working hypothesis, suppose the mass density is scaled as $ \rho (r) \to \rho (r / a) $, that is, in Fig.\ref {f1} left, the graph simply shifts horizontally when looking at different galaxies. It is easy to verify that all structural elements that define the position of key points in the scenario withstand this scaling. The mass function shown in Fig.\ref {f1} right scales as $ M_ {grav} (R) \to4 \pi \int_0 ^ R dr r ^ 2 \rho_ {grav} (r/a) = a ^ 3M_ {grav} (R / a) $, the $ M_ {vac} (R) $ contribution is also scaled, which must be added here in cosmological estimates. Thus, the total mass of dark matter in the galaxy under the taken assumptions is scaled as $ M_ {dm, gal} \to M_ {dm, gal} a ^ 3 $. At the same time, the square of the orbital velocity $ v ^ 2 = GM / R $ is scaled as $ v ^ 2 \to v ^ 2a ^ 2 $, the velocity is scaled as $ v \to va $. Due to the Tully-Fisher relation, the luminous mass scales as $ M_ {lm} \to M_ {lm} a ^ 4 $. Thus, $ M_ {dm} \sim (M_ {lm}) ^ {3/4} $ proportionality holds along the sequence under consideration, the required correction factor is $ N_ {gal} \eta_ {3/4} = 8.302 \cdot10 ^ { 10} $.

Note that the function $ N_ {gal} \eta_p $ has a minimum at $ p \sim0.9 $ and changes little in the range $ p = 0.3 ... 2 $, so other dependencies $ M_ {dm} \sim (M_ { lm}) ^ p $ lead to a similar result for $ p $ in this interval. Also note that in other works, other values of $ p $ were obtained, \cite {astro-ph/9304018} $ p = 0.3 $, \cite {Girardi_2002_ApJ_569_720} $ p = 1.34 $, \cite {astro-ph/0703115 } Eq. (7) $ p = 0.3-1.1 $ for spiral galaxies, \cite {1609.06903} Eq. (21) $ p = 1.05-1.24 $ for dwarf disc galaxies. This result strongly depends on the choice of the mass profile and the halo cutoff radius. In our scenario S1.3, the cutoff is applied at the outer radius $ R_ {cut2} $, where the phase transition of dark matter into dark energy is completed, outside of which the density of dark matter vanishes.

Compared with the value obtained in S1.3 required for joining the relations $ N_ {gal} \eta_ {3/4} = \sum M_ {dm, gal} / M_ {dm, MW} \sim8 \cdot10 ^ {10} $ and $ M_ {dm, uni} / M_ {dm, MW} \sim9 \cdot10 ^ {11} $, there is a discrepancy of 1.1dex. If we change the modeling of S1.3 a little and achieve an exact fit to the experimental estimate \cite {1405.0306} $ M_ {MW} \sim 8 \cdot10 ^ {11} M_ \odot $, we get a discrepancy of 0.8dex. Thus, our assumption about the scale invariance of scenario S1.3 fits into the existing scatter of experimental data. At the same time, it becomes clear that the remaining discrepancy, in fact, is not related to the details of our modeling, but is the result of direct comparison of different experimental estimates. Using relations of the form $ M_ {dm} \sim (M_ {lm}) ^ p $ from the experimental works cited above, a similar result will be obtained.

A similar result will also be obtained in our other scenarios if we accept the same scaling assumptions in them, that is, $ R_ {cut} \sim a $ and $ \epsilon \sim a ^ 2 $. In this case, the contributions $ M_ {dm, gal} $ and $ M_ {vac} $ are scaled in the same way $ \sim a ^ 3 $, and if for the initially taken MW galaxy the contribution $ M_ {vac} $ can be neglected, then it can be neglected along the entire sequence. The relation $ N '_ {gal} \sim N_ {gal} \eta_ {3/4} $ is subject to verification, where $ N' _ {gal} = 1.7 \cdot10 ^ {12} -1.4 \cdot10 ^ {11} $ for $ R_ {cut} =$50kpc-0.6Mpc. The deviation from $ N_ {gal} \eta_ {3/4} = 8.302 \cdot10 ^ {10} $ here is 1.3dex-0.2dex, with a preference for larger values of $ R_ {cut} $.

\paragraph * {Structure of solutions of the nonlinear Klein-Gordon equation.} In general, solutions of this equation for stationary spherically symmetric problems do not admit separation of variables and cannot be represented in the form $ \phi (t, r) = e ^ {iEt} s ( r) $ or their linear combinations. However, in the special case of harmonic boundary conditions $ \phi (t, r_1) = e ^ {iEt} s_1 $, $ \phi'_r (t, r_1) = e ^ {iEt} d_1 $, the following consideration can be used. Let's perform the numerical integration of this equation using the finite difference scheme 
\begin{eqnarray}
&&\df^2\phi/\df t^2=(\phi(t+dt,r)+\phi(t-dt,r)-2\phi(t,r))/dt^2,\\
&&\df^2\phi/\df r^2=(\phi(t,r+dr)+\phi(t,r-dr)-2\phi(t,r))/dr^2,\\
&&\df\phi/\df r=(\phi(t,r+dr)-\phi(t,r-dr))/(2dr). 
\end{eqnarray}
This is not the scheme that is used in practice to solve such equations, but here, for the purpose of proof, it can be applied with a sufficiently small choice of integration steps. When integrating the solution by layers of constant $ r $, starting from $ r = r_1 $ and then recurrently into the region $ r <r_1 $, it is easy to check that the equation solved with respect to the leading layer $ \phi (t, r-dr) $ will always carry the phase factor $ e ^ {iEt} $ and the amplitude $ s (r) $ depending only on $ r $. Thus, the solution will globally have the form $ \phi (t, r) = e ^ {iEt} s (r) $. In the particular case $ E = 0 $, the solution will be globally stationary $ \phi (t, r) = s (r) $.

A slight subtlety is that this reasoning is valid for finite $ r_1 $, while for the considered solutions the boundary condition is imposed for $ r \to \infty $, asymptotically. This problem can be solved as follows. Let's approximate the potential near the minimum by an analytically solvable form (\ref {V1q}). The solutions will be (\ref {V1sol}) with $ C_2 = 0 $. For a finite value of $ r_1 $, when the solution is still in the considered vicinity of the minimum, calculate $ s (r_1) $ and $ s' (r_1) $ on the solution and use them as boundary conditions in the above numerical integration scheme. As a result, a global solution of the required form will be obtained.

Note also that as long as the consideration concerns stationary solutions, there is no difference between the theories of complex and real scalar field. The difference appears for $ E \neq0 $, the real harmonic solutions have the form $ \phi = \cos (Et) s (r) $. In this case, the time dependence penetrates into the argument of the potential function, and there is no reduction in the dimension. In the nonlinear theory of real scalar field, single-frequency harmonic functions can only be approximate solutions \cite {1710.08910}.

\paragraph * {NRDM|TOV system.} Stationary spherically symmetric gravitational fields are described by the metric tensor and the energy-momentum tensor 
\begin{eqnarray}
&g_{\mu\nu}=\diag(-A(r),B(r),r^2,(r\sin\theta)^2),\ T^\mu_\nu=\diag(-\rho(r),p_r(r),p_t(r),p_t(r)),
\end{eqnarray}
in the coordinate system composed of the time of distant observer and the standard spherical coordinate system $ x ^ \mu = (t, r, \theta, \phi) $. Einstein's equations for such problems are: 
\begin{eqnarray}
&\rho=(-B + B^2 + r B'_r)/(8 \pi r^2 B^2), \ p_r=(A - A B + r A'_r)/(8 \pi r^2 A B),\\
&p_t=(-r B (A'_r)^2 - 2 A^2 B'_r + A (-r A'_r B'_r + 2 B (A'_r + r A''_{rr})))/(32 \pi r A^2 B^2).
\end{eqnarray}
This system can be taken from \cite{Visser1996} (11.36-38) and converted to our notations or derived from the first principles using Mathematica code \cite{Hartle}. It is easy to check that this system satisfies the relation 
\begin{eqnarray}
&r(p_r+\rho)A'_r+2A(r(p_r)'_r+2p_r-2p_t)=0. \label{hydroeq}
\end{eqnarray}
This equation has a profound meaning, as a consistency condition for the Einstein equations, Bianchi identity. It is equivalent to conservation of energy-momentum $\nabla_\mu T^\mu_\nu=0$. It also has a physical meaning of hydrostatic equation, since it describes a distribution of pressure and density in a stationary spherically symmetric gravitational field.

Now let's consider EOS for two phases
\begin{eqnarray}
&\textrm{NRDM:}\ p_r=\rho,\ p_t=0,\quad \textrm{TOV:}\ p_r=p_t=w\rho,\ w=1/3
\end{eqnarray}
and join the corresponding solutions of the Einstein equations. First of all, the analysis of the matching conditions for the hydrostatic equation shows that at the phase boundary $ p_r $ must be $ C ^ 0 $-continuous, while $ \rho $ and $ p_t $ undergo a jump, repeating the jump of EOS. Further, analyzing the behavior of the metric profiles in the Einstein equations, we see that $ A $ must be $ C ^ 1 $-continuous, and $ B $ - $ C ^ 0 $-continuous.

The hydrostatic equation for such EOS can be solved analytically: 
\begin{eqnarray}
&\textrm{NRDM:}\ p_r=\epsilon/(8\pi r^2A),\quad \textrm{TOV:}\ p_r=k/(4\pi A^2), \label{hydreqsol}
\end{eqnarray}
where $ \epsilon $ and $ k $ are integration constants. Further, the Einstein equations for our combined model are: 
\begin{eqnarray}
&\textrm{NRDM:}\ a'_x=-1+e^b+\epsilon\, e^{b-a},\ b'_x=1-e^b+\epsilon\, e^{b-a},\\
&\textrm{TOV:}\ a'_x=-1 + e^b + 2k e^{2 x -2a + b},\ b'_x= 1 - e^b + 6k e^{2 x -2a + b},
\label{tov1}
\end{eqnarray}
in logarithmic variables $ x = \log r $, $ a = \log A $, $ b = \log B $. Integration starts in the NRDM phase from a point distant from the center, where the initial conditions are selected $ a_1 = 0 $, $ b_1 = - \log (1-2M_1 / r_1) $, $ 2M_1 = \epsilon r_1 + r_s $. In this case, $ C ^ 0 $-continuous matching of $ p_r $ at the phase boundary $ r = r_ {2 *} $ leads to the condition $ k = \epsilon A_ {2 *} / (2r_ {2 *} ^ 2) $, whereby the equation (\ref {tov1}) can be rewritten as
\begin{eqnarray}
&\textrm{TOV:}\ a'_x=-1 + e^b + \epsilon\, e^{2 (x-x_{2*}) -2a + b +a_{2*}},\\
&b'_x= 1 - e^b + 3\epsilon\, e^{2 (x-x_{2*}) -2a + b +a_{2*}}.
\end{eqnarray}
From this it is clear that $ C ^ 0 $-continuous matching of $ (a, b) $ at the phase boundary enables $ C ^ 0 $-continuous matching also for the derivative $ a'_x $, while $ b'_x $ undergoes a jump by a factor 3. This provides the continuity mode $ C ^ 1 $ for $ a $, $ C ^ 0 $ for $ b $, in accordance with the above theoretical analysis.

\end{document}